\documentclass[12pt,preprint]{aastex}
\shorttitle{The UV-Optical Galaxy Color-Magnitude Diagram I}
\shortauthors{Wyder et al.}
\slugcomment{Submitted for publication in the Special {\it GALEX} Ap.J.Suppl. Issue}
\begin{document}
\title{The UV-Optical Galaxy Color-Magnitude Diagram I: Basic Properties}

\author{Ted K. Wyder\altaffilmark{1,2},
D. Christopher Martin\altaffilmark{1},
David Schiminovich\altaffilmark{3},
Mark Seibert\altaffilmark{4},
Tam\'as Budav\'ari\altaffilmark{5},
Marie A. Treyer\altaffilmark{1,6},
Tom A. Barlow\altaffilmark{1},
Karl Forster\altaffilmark{1},
Peter G. Friedman\altaffilmark{1},
Patrick Morrissey\altaffilmark{1},
Susan G. Neff\altaffilmark{7},
Todd Small\altaffilmark{1},
Luciana Bianchi\altaffilmark{5},
Jos\'e Donas\altaffilmark{6},
Timothy M. Heckman\altaffilmark{8},
Young-Wook Lee\altaffilmark{9},
Barry F. Madore\altaffilmark{4},
Bruno Milliard\altaffilmark{6},
R. Michael Rich\altaffilmark{10},
Alex S. Szalay\altaffilmark{8},
Barry Y. Welsh\altaffilmark{11}, 
Sukyoung K. Yi\altaffilmark{9}
}

\altaffiltext{1}{California Institute of Technology, MC 405-47, 1200 East
California Boulevard, Pasadena, CA 91125}

\altaffiltext{2}{e-mail: wyder@srl.caltech.edu}

\altaffiltext{3}{Department of Astronomy, Columbia University, New York, NY 10027}

\altaffiltext{4}{Observatories of the Carnegie Institution of Washington,
813 Santa Barbara St., Pasadena, CA 91101}

\altaffiltext{5}{Center for Astrophysical Sciences, The Johns Hopkins
University, 3400 N. Charles St., Baltimore, MD 21218}

\altaffiltext{6}{Laboratoire d'Astrophysique de Marseille, BP 8, Traverse
du Siphon, 13376 Marseille Cedex 12, France}

\altaffiltext{7}{Laboratory for Astronomy and Solar Physics, NASA Goddard
Space Flight Center, Greenbelt, MD 20771}

\altaffiltext{8}{Department of Physics and Astronomy, The Johns Hopkins
University, Homewood Campus, Baltimore, MD 21218}

\altaffiltext{9}{Center for Space Astrophysics, Yonsei University, Seoul
120-749, Korea}

\altaffiltext{10}{Department of Physics and Astronomy, University of
California, Los Angeles, CA 90095}

\altaffiltext{11}{Space Sciences Laboratory, University of California at
Berkeley, 601 Campbell Hall, Berkeley, CA 94720}

\begin{abstract}

We have analyzed the bivariate distribution of galaxies as a function of ultraviolet-optical colors and absolute magnitudes in the local universe. The sample consists of galaxies with redshifts and optical photometry from the Sloan Digital Sky Survey (SDSS) main galaxy sample matched with detections in the near-ultraviolet ($NUV$) and far-ultraviolet ($FUV$) bands in the Medium Imaging Survey being carried out by the {\it Galaxy Evolution Explorer (GALEX)} satellite. In the $(NUV-r)_{0.1}$ vs. $M_{r,0.1}$ galaxy color-magnitude diagram, the galaxies separate into two well-defined blue and red sequences. The $(NUV-r)_{0.1}$ color distribution at each $M_{r,0.1}$ is not well fit by the sum of two Gaussians due to an excess of galaxies in between the two sequences. The peaks of both sequences become redder with increasing luminosity with a distinct blue peak visible up to $M_{r,0.1}\sim-23$. The $r_{0.1}$-band luminosity functions vary systematically with color, with the faint end slope and characteristic luminosity gradually increasing with color. After correcting for attenuation due to dust, we find that approximately one quarter of the color variation along the blue sequence is due to dust with the remainder due to star formation history and metallicity. Finally, we present the distribution of galaxies as a function of specific star formation rate and stellar mass. The specific star formation rates imply that galaxies along the blue sequence progress from low mass galaxies with star formation rates that increase somewhat with time to more massive galaxies with a more or less constant star formation rate. Above a stellar mass of $\sim 10^{10.5}$ M$_{\sun}$, galaxies with low ratios of current to past averaged star formation rate begin to dominate.

\end{abstract}


\section{Introduction}

Galaxies exhibit bimodal distributions in a number of observed properties. The bimodality in galaxy morphologies formed the basis of the original galaxy classification scheme of \citet{hubble26}. The colors and luminosities of galaxies have been long known to correlate with morphology \citep[e.g.][]{devaucouleurs61, chester64} with ellipticals being predominantly red and spirals and irregulars blue. 

More recently, large statistical samples of galaxies have become available, allowing us to investigate the bimodality of galaxies in a much more quantitative way.  In particular, the bimodality appears quite strongly in the galaxy $(u-r)$ color distribution which consists of two peaks with a minimum in between them at $(u-r)\approx 2.1-2.2$ \citep{strateva01}. Galaxies in the red peak tend to be predominantly morphologically early-type and high surface brightness galaxies while those in the blue peak are dominated by morphologically late-type galaxies with lower surface brightness \citep{strateva01,driver06,blanton03b,ball06}. Based upon a sample of low-redshift galaxies from the SDSS, \citet{baldry04} investigated the distribution of galaxies in the $(u-r)$ vs. $M_r$ color-magnitude diagram (CMD). The galaxies in their sample separate into blue and red sequences with the distribution in color at each absolute magnitude well-fit by the sum of two Gaussians. The mean color as a function of $M_r$ for each sequence consists of an overall reddening with increasing luminosity with a steeper transition in the average color and width of both sequences at a stellar mass of $\sim2\times10^{10}$ M$_{\sun}$. 

In addition to mass, one of the most important other factors suspected of contributing to the galaxy bimodality is the environment. While it has long been known that the morphologies of galaxies are correlated with the local density \citep{dressler80}, the dependence of galaxy colors and luminosities with local density is complicated. Although the ratio of the number of red to blue galaxies varies strongly with the local density, the mean color of the blue and red sequences varies relatively little with environment \citep{balogh04}. On the other hand, the luminosity of blue sequence galaxies is nearly independent of environment while both luminous and faint red galaxies are found on average in higher density environments than intermediate luminosity red galaxies \citep{hogg03}.

The galaxy bimodality has also begun to be investigated based upon large samples of galaxy spectra from the Sloan Digital Sky Survey (SDSS).  In particular, \citet{kauffmann03a} developed a method that uses the Balmer absorption line index ${\rm H\delta_A}$ and the ${\rm 4000 \AA}$ break strength $D_n(4000)$ measured from the SDSS fiber spectra in the central $3\arcsec$ of each galaxy to constrain the star formation histories, dust attenuation, and stellar masses for their sample. Based upon these derived parameters, \citet{kauffmann03b} showed that galaxies tend to divide into two distinct groups around a stellar mass of $3\times10^{10}$ M$_{\sun}$, similar to the transition mass noted in the optical galaxy CMD \citep{baldry04}. While galaxies below this mass tend to have younger stellar populations, more massive galaxies tend to be older. In related work, \citet{brinchmann04} used the emission lines in the SDSS spectra to determine star formation rates (SFRs) for a large sample of SDSS galaxies. Using the specific star formation rate, i.e. the current SFR divided by the stellar mass $M_*$, \citet{brinchmann04} found that galaxies with $10^8<M<10^9$ M$_{\sun}$ have $\log{(SFR/M^*)}=-9.6$ to $-10$, values consistent with an approximately constant SFR with time. Above $10^{10}$ M$_{\sun}$, the specific SFRs decline with mass, implying star formation histories increasingly weighted to much older ages.

The evolution of the galaxy color-magnitude diagram out to $z\sim1$ has begun to be explored \citep{willmer06, faber06, blanton06}. These results show that the galaxy bimodality is already in place at $z\sim1$. However, the color of both sequences tends to become somewhat bluer with increasing redshift while the luminosity function of both red and blue galaxies shifts to higher luminosities \citep{blanton06, willmer06}. Based upon combining the DEEP2 and COMBO17 surveys, \citet{faber06} argued that the number density of blue galaxies is more or less constant from $z\sim1$ to $z\sim0$, while the number density of red galaxies has been increasing. \citet{faber06} proposed a scenario to explain their data in which some blue galaxies migrate to the red sequence as a result of gas-rich mergers that use up the remaining gas in an interaction-induced starburst. These galaxies then migrate up the red sequence by a series of gas-free mergers.

The origin of the galaxy bimodality and corresponding transition mass of a few$\times10^{10}$ M$_{\sun}$ is beginning to be understood theoretically. Based upon a semi-analytic model utilizing some simple prescriptions for gas cooling, star formation, and supernova feedback coupled with merging histories of dark matter haloes, \citet{menci05} modeled the $(u-r)$ vs. $M_r$ CMD of \citet{baldry04}. In their model, feedback from supernovae is ineffective at regulating star formation for galaxies above a certain threshold halo mass. In these massive galaxies all of the gas is consumed relatively quickly and results in a red sequence galaxy at zero redshift. Blue sequence galaxies, on the other hand, tend to come from less massive progenitors where supernovae feedback is effective at regulating star formation, thus allowing star formation to continue down to the present. While their model is successful at reproducing most of the optical CMD, it predicts too many blue galaxies at $M_r=-22$ compared to the observations.

A different explanation for the origin of bimodality has been suggested by \citet{dekel06}. According to this model, above a critical halo mass $M_{shock} \sim 10^{12}$ M$_{\sun}$, a shock is generated in the gas accreting onto the dark matter halo which heats most of the gas and prevents it from cooling and forming stars. In these massive haloes, star formation does happen at $z\gtrsim2$ due to cold gas that is able to penetrate the hot gas, leading to a burst of star formation, while for $z\lesssim2$ heating from Active Galactic Nuclei (AGN) prevents gas from forming any more stars. This naturally leads to the most massive galaxies lying on the red sequence at $z\sim0$. For galaxies residing in halos with masses less than $10^{12}$ M$_{\sun}$, the gas is not shock heated, allowing cold flows to fuel star formation that is then regulated by supernova feedback. As a result, lower mass galaxies lie on the blue sequence and the location of the bright tip of the blue sequence is due to the onset of the shock in the accreting gas for more massive halos and the feedback from AGN. In this scenario, galaxies tend to move up the blue sequence with time until their masses go above $M_{shock}$, or they merge into another more massive halo with mass above $M_{shock}$, after which the gas in the galaxy is no longer allowed to cool and star formation ceases. Both \citet{cattaneo06} and \citet{croton06} have coupled semi-analytic models including the transition from shock heating to cold flows and feedback from supernova as well as AGN with the merging histories of dark matter haloes from N-body simulations. While the details of the modeling of the baryonic physics differs somewhat, both groups were able to reproduce the local galaxy CMD by tuning the various parameters affecting star formation and feedback in their models.

In this paper, we investigate the galaxy bimodality as revealed in the UV minus optical colors of a large sample of galaxies observed by both the {\it Galaxy Evolution Explorer} ({\it GALEX}) and the SDSS. While significant contributions to the UV luminosity can come from older evolved stars in red sequence galaxies \citep[e.g.][]{yi05, rich05}, in general the UV light in galaxies is dominated by massive stars with main sequence lifetimes up to $\sim 10^8$ yrs. As a result, the emerging UV luminosity is proportional to the recent star formation rate once corrected for light absorbed by dust \citep{kennicutt98}. The greater sensitivity of the {\it GALEX} bands to the recent star formation rate as compared to the SDSS $u$ band would lead us to expect a greater separation between the red and blue sequences. While the measurements from the SDSS spectra are sensitive diagnostics of the stellar populations in galaxies, they are measured only in the central $3\arcsec$ of each galaxy, making somewhat uncertain aperture corrections necessary to account for the portion of each galaxy not sampled. While the UV data presented here is much more susceptible to dust attenuation than in the optical SDSS data, the UV measurements sample the entire galaxy and thus complement the SDSS measurements. 

\section{Data and Analysis}

\subsection{{\it GALEX} Data}

The UV data presented here are derived from the {\it GALEX} Medium Imaging Survey (MIS) \citep{martin05}. {\it GALEX} is a 50cm diameter UV telescope that images the sky simultaneously in both a $FUV$ and a $NUV$ band, centered at 1540 \AA~and 2300 \AA, respectively. The field-of-view of {\it GALEX} is approximately circular with a diameter of $1\fdg2$ and resolution of about $5\farcs5$ FWHM in the $NUV$. The MIS pointings are chosen to overlap areas of sky with imaging and spectroscopy from the SDSS and consist of exposures of at least one to a few orbits with the mode of the exposure time distribution being 1700 sec. The dataset used in our analysis is taken from the union of the {\it GALEX} first data release (GR1) with the {\it GALEX} internal release IR1.1, a subset of which has been included in the second data release (GR2) publicly available from the {\it GALEX} archive.\footnote{The {\it GALEX} archive can be accessed from http://galex.stsci.edu/GR2/.} The IR1.1 data was processed with a pipeline very similar to that used in the GR1 data and employed the same calibration as used in that release. Details of the {\it GALEX} detectors, pipeline, calibration and source extraction can be found in \citet{morrissey05,morrissey07}. 

The {\it GALEX} pipeline uses the SExtractor program \citep{bertin96} to detect and make measurements of sources in the images. Throughout this paper, we use the "MAG\_AUTO" measurements output by SExtractor. These magnitudes are measured within elliptical apertures with semi-major axis scaled to 2.5 times the first moment of each source's radial profile, as first suggested by \citet{kron80}. 

Due to an error in the way in which the SExtractor parameters were set in the GALEX pipeline, the photometric errors for most of the sources, as reported in the GR1 and IR1.1 catalogs, are underestimated. For each source we have calculated a more accurate statistical error in the total magnitude from the size of the MAG\_AUTO aperture and the flat field response, exposure time, and sky background at the source position. In addition to the statistical errors, we have added in quadrature an assumed zero-point plus flat field uncertainty of 2\% in the $NUV$ and 5\% in the $FUV$ \citep{morrissey07}. The errors increase from the zeropoint uncertainty at the bright end up to $\approx 0.2-0.3$ mag at 23rd mag in both bands.

\subsection{{\it GALEX}-SDSS Matched Sample}

The {\it GALEX} MIS catalogs were matched with the SDSS MPA/JHU DR4 value-added catalogs.\footnote{The MPA/JHU value-added catalogs were downloaded from http://www.mps-garching.mpg.de/SDSS/.} These catalogs consist of line and index measurements from the SDSS spectra as well as many derived quantities and are described in more detail in a series of papers on the star formation rates, star formation histories, stellar masses, and metallicities of galaxies in the local universe \citep{kauffmann03a,kauffmann03b,brinchmann04,tremonti04}. For each {\it GALEX} pointing, SDSS sources within $0\fdg6$ of the {\it GALEX} field center were matched with the nearest {\it GALEX} source within a radius of $4\arcsec$. When concatenating together the catalogs for all the fields, we removed duplicate {\it GALEX} detections in the overlap regions between adjacent pointings by using the SDSS identification numbers (Plate ID, MJD, Fiber ID) and selecting the {\it GALEX} match closest to its field center. 

After matching the {\it GALEX} and SDSS data, we further restricted the sample with various cuts intended to generate a complete statistical sample which are summarized in Table \ref{selection_limits}. For the SDSS photometry, we selected galaxies targetted for spectroscopy in the SDSS main galaxy sample with $r$-band magnitudes in the range $14.5 < r < 17.6$.  While the nominal magnitude limit of the SDSS main galaxy sample is $r=17.77$ \citep{strauss02}, in practice the limit varies as a function of position on the sky. After examining the galaxy number counts, we set the faint limit to $r=17.6$ because the counts begin to turn-over below this level. While the median photometric error for the SDSS galaxies is only 0.03 mag, there are a small fraction with much larger errors. In order to remove these objects, we restricted the sample to galaxies with errors $\sigma_r < 0.2$ mag. In addition to the photometry, we further restricted the sample to those galaxies with redshifts $z$ in the range $0.01 < z  < 0.25$ and with redshift confidence $z_{conf} > 0.67$. 

In addition to the cuts on the SDSS data, we applied several cuts based upon the UV measurements. Since {\it GALEX} photometry and astrometry degrade near the edge of the detectors, we only included objects in the sample if their distance from the {\it GALEX} field center $fov\_radius$ is less than 0\fdg55. {\it GALEX} detections were also required to have $nuv\_artifact\leq1$. This excludes from our sample galaxies that lie within regions expected to be contaminated by reflections from bright stars within the field. Areas of the images with $nuv\_artifact=1$ are designed to flag regions where scattered light is predicted from bright stars just outside the field-of-view. Currently, this flag is set very conservatively and the vast majority of sources with this flag set are in regions that are free from scattered light issues. We therefore elected to ignore this flag for our sample. We also only included fields with exposure times greater than 1000 sec. The resulting overlap area between {\it GALEX} and the SDSS including all of the above cuts is 485.321 deg$^2$ in the $NUV$ and 411.266 deg$^2$ in the $FUV$. The $FUV$ sample has a somewhat smaller area because some of the fields have $NUV$ data only. The procedure used to calculate the overlap area is similar to that used by \citet{bianchi07}, where a more detailed description can be found.

\begin{figure}
\includegraphics*[width=5in,height=5in]{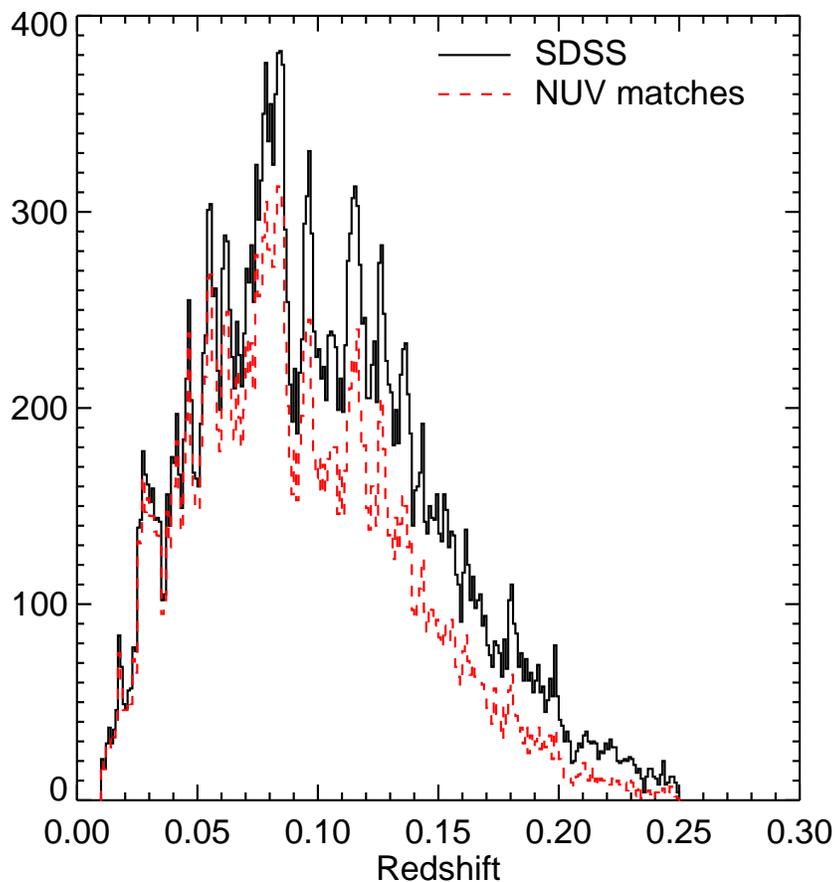}
\caption{The redshift distributions for the $NUV$ sample. The solid black line indicates the redshift distribution for the SDSS main galaxy sample covered by the {\it GALEX} fields with $NUV$ exposure times greater than 1000 sec and satisfying the SDSS selection criteria in Table \ref{selection_limits}. The dashed red line is the redshift distribution for galaxies with a {\it GALEX} match within $4\arcsec$ and satisfying the {\it GALEX} plus SDSS selection criteria in Table \ref{selection_limits}.
\label{zdist_nuv}}
\end{figure}

\begin{figure}
\includegraphics*[width=5in,height=5in]{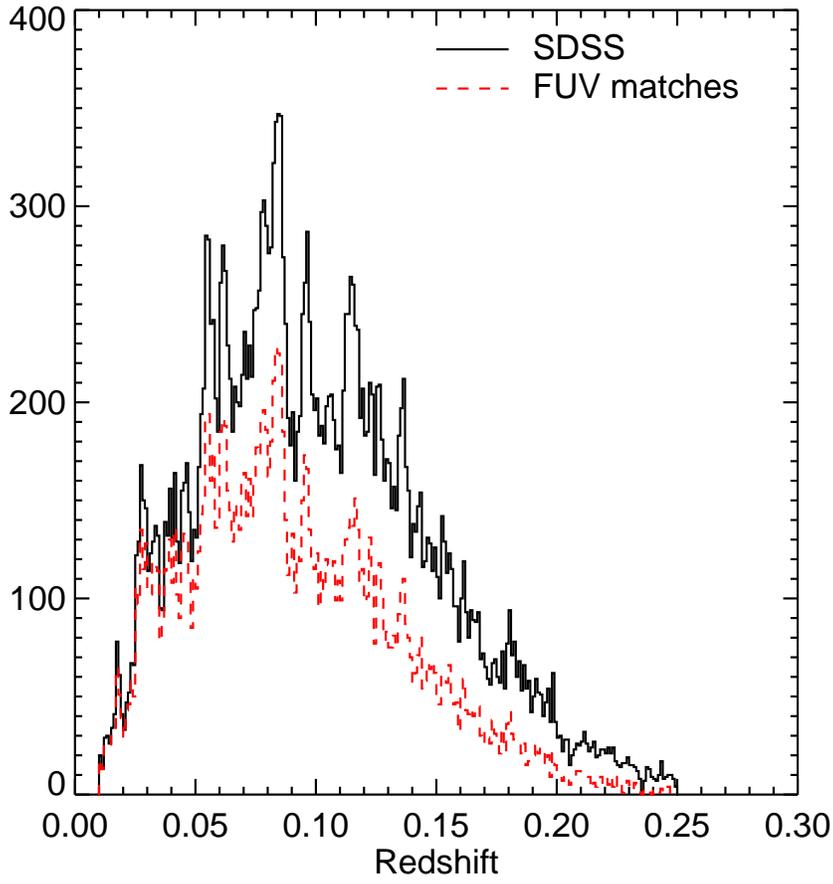}
\caption{The same as in Figure \ref{zdist_nuv} except for the $FUV$ sample.\label{zdist_fuv}}
\end{figure}

\begin{figure}
\includegraphics*[width=5in,height=5in]{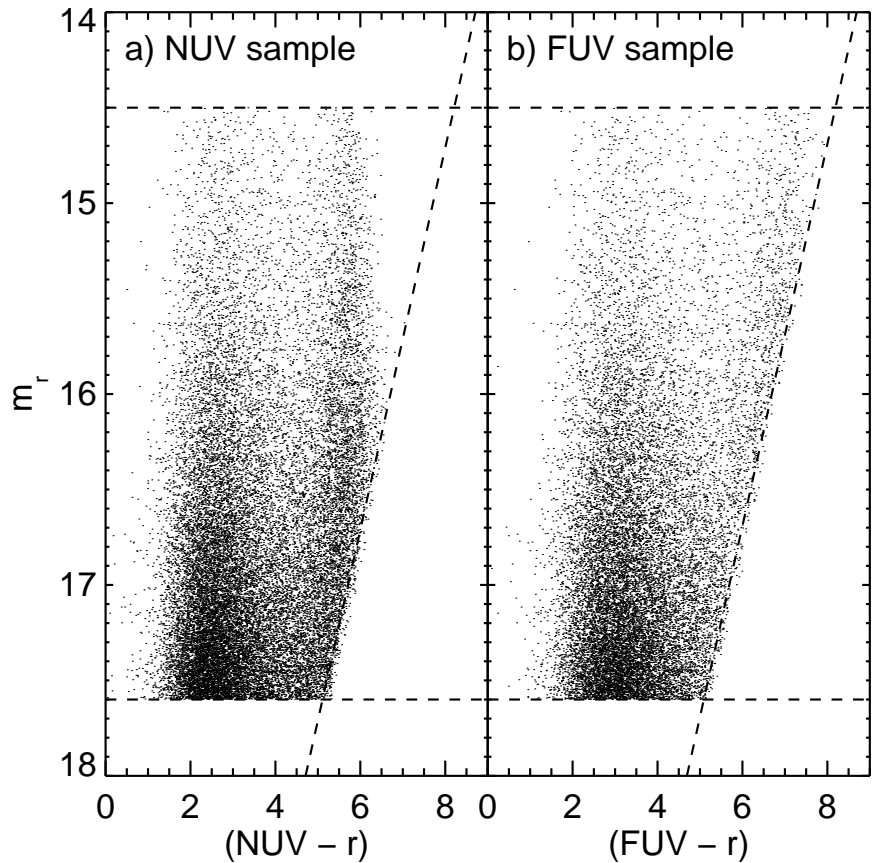}
\caption{The apparent $r$-band magnitude as a function of a) $(NUV-r)$ color for the $NUV$-selected sample and b) $(FUV-r)$ color for the $FUV$-selected sample. In both panels the horizontal dashed lines indicate the $r$-band bright and faint magnitude limits while  the diagonal dashed lines indicate the $FUV$ and $NUV$ magnitude limit of 23, corrected to the median Galactic extinction in the samples. In the $NUV$ sample, the lack of galaxies in the upper right hand corner would indicate that the fall-off in galaxies at the red end of the color distribution is real and not due to the UV magnitude limit. This does not appear to be the case for the $FUV$ sample, where the data go all the way up to the magnitude limit. In this case, we may not be sampling the full color distribution of galaxies.
\label{apparent_cmd}}
\end{figure}

For both the $FUV$ and $NUV$ samples, we included galaxies with apparent magnitudes in the range $16 < N(F)UV < 23$. Based upon a series of artificial source tests, we have estimated that the data is greater than 80\% complete for $NUV, FUV < 23$ mag.  After applying the Galactic extinction correction, the UV magnitude limit varies across the sky. We account for this variation below when computing the volume densities.  After applying all of the above cuts to the data, the $NUV$ and $FUV$ samples contain 26,281 and 18,091 galaxies, respectively. The redshift distributions for the $NUV$ and $FUV$ samples are shown in Figures \ref{zdist_nuv} and \ref{zdist_fuv}. In the figures, the solid black lines are the distributions of SDSS galaxies lying within the area observed by {\it GALEX} and satisfying the SDSS selection criteria while the dashed red lines show those galaxies which have a {\it GALEX} match falling within the {\it GALEX} selection criteria as well. The fraction of SDSS main sample galaxies with {\it GALEX} matches tends to decline with redshift, mainly due to increasing numbers of red galaxies falling below the {\it GALEX} detection limit. In Figure \ref{apparent_cmd} the $r$-band magnitude is plotted as a function of color for the $NUV$ and $FUV$ samples. The dashed lines in the figure indicate the $r$-band and UV magnitude limits. For the $NUV$ sample, the lack of galaxies with $(NUV-r) \gtrsim 6.5$ at bright $r$ magnitudes would argue that the {\it GALEX} MIS data is deep enough to fully sample the entire color distribution of galaxies in the local universe. Thus, the edge of the color distribution is likely real and not a selection effect. On the other hand, for the $FUV$ sample in the right hand panel of Figure \ref{apparent_cmd}, the data go right up to the selection limit at the red end. In the $FUV$ sample, the red edge of the color distribution thus reflects the UV magnitude limit.

\begin{figure}
\includegraphics*[width=5in,height=5in]{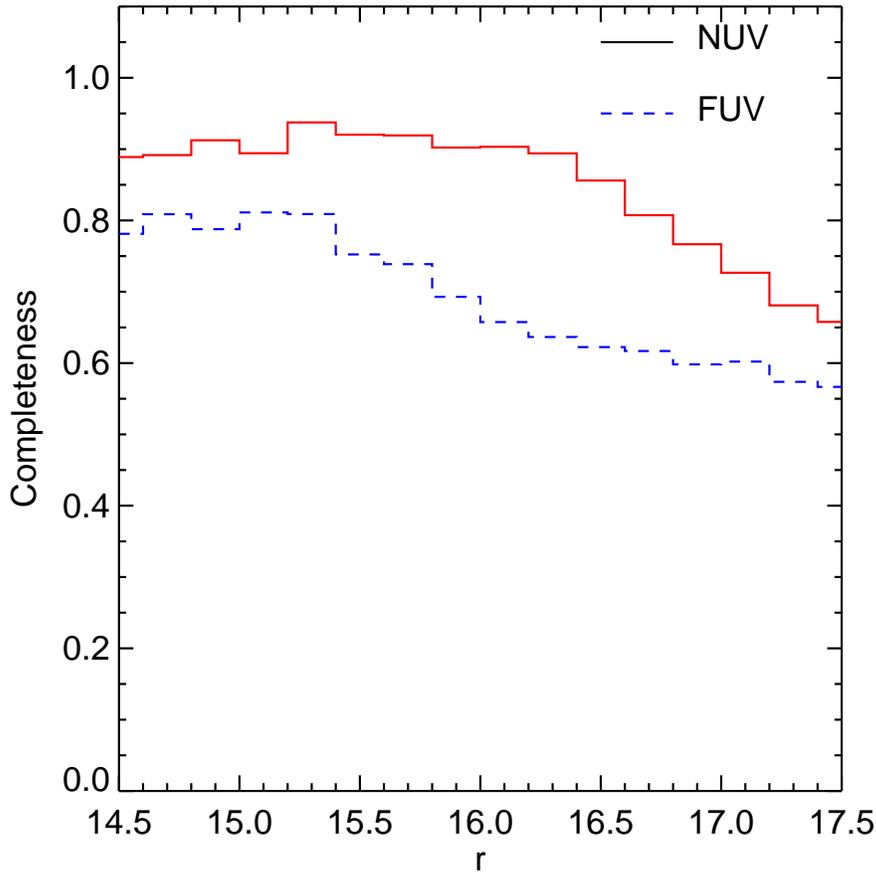}
\caption{The fraction of galaxies in the SDSS main galaxy sample with {\it GALEX} $NUV$ (red solid line) and $FUV$ (blue dashed line) matches as a function of $r$ magnitude. The fall in the $NUV$ completeness for $r \lesssim16.5$ is due to the loss of red galaxies below the $NUV$ magnitude limit. The completeness at the bright end is about 90\%. While most of these galaxies are visible in the {\it GALEX} images, they are not matched to the SDSS sample due to shredding of large galaxies in the {\it GALEX} data or due to the UV centroid being offset from the SDSS position by more than the $4\arcsec$ search radius used to match the two catalogs. The fall-off in the $FUV$ completeness occurs at $r>15.5$ due to the larger spread in $(FUV-r)$ colors compared to the $(NUV-r)$ distribution.\label{completeness_r}}
\end{figure}

The fraction of SDSS main sample galaxies that lie within 0\fdg55 of a {\it GALEX} MIS field center and that have a {\it GALEX} match within our $4\arcsec$ search radius is shown in Figure \ref{completeness_r} as a function of $r$ magnitude. In the $NUV$ sample, the completeness is roughly constant at about 90\% down to $r\approx16.5$. Fainter than this magnitude, the completeness begins to drop off. There are very few galaxies with colors redder than $(NUV-r)\approx6.5$. At the {\it GALEX} magnitude limit of 23 mag, a galaxy with this red a color would have $r=16.5$. Thus, the fraction of SDSS galaxies with a {\it GALEX} match begins to drop fainter than this $r$ magnitude due to increasing numbers of red galaxies falling below the {\it GALEX} detection limit. For the $FUV$ sample, the reddest galaxies have $(FUV-r)\approx7.5$ which corresponds to $r=15.5$ at the limiting magnitude of $FUV=23$. As expected, the fraction of SDSS galaxies with an $FUV$ match begins to decline at about this $r$ magnitude.

In both the $FUV$ and $NUV$ samples, the match completeness does not reach 100\% at bright $r$ magnitudes. We have visually inspected all SDSS main sample galaxies with $14.5 < r < 15.5$ that were observed by {\it GALEX}. For almost all of these galaxies, there is a galaxy visible in the {\it GALEX} images. However, the UV center measured by the {\it GALEX} pipeline for these non-matches lies more than $4\arcsec$ from the SDSS position. Sometimes the {\it GALEX} pipeline breaks the galaxy into more than one fragment, none of which are coincident with the SDSS position. In other cases, especially if the galaxy has a low UV surface brightness, the center can be offset from the SDSS position by more than our search radius even if the {\it GALEX} pipeline detects the galaxy as a single object. We assume that the level portion of the match completeness curves in Figure \ref{completeness_r} gives the intrinsic completeness for {\it GALEX} detections of the SDSS main galaxy sample. The values we adopt are 0.91 and 0.80 for the $NUV$ and $FUV$, respectively.

While the completeness of the SDSS photometric sample is nearly 100\%, some fraction of galaxies that meet the SDSS main galaxy sample selection do not have a redshift measured \citep{strauss02}. Although some galaxies do not have a redshift due to low signal-to-noise in their spectra, the majority of targeted galaxies without redshifts  are missed due to the constraint that SDSS spectroscopy fibers can not be placed closer than $55\arcsec$ to one another. Some of these missed galaxies can be observed in neighboring plates if that region of sky is covered by more than one plate. While the exact completeness is determined by the precise geometry of the spectroscopic plates, the result is that the spectroscopic completeness of the SDSS main galaxy sample is 92-94\% for the early release data \citep{strauss02,blanton01}. We have adopted a spectroscopic completeness of 0.9. Multiplying the {\it GALEX}-SDSS match completeness by the SDSS spectroscopic completeness, we estimated the total completeness of our sample to be 0.82 in the $NUV$ and $0.72$ in the $FUV$. In calculating the volume densities below, we correct the number counts by these factors, (i.e. the factor $f$ in equations (\ref{phi_eqn}) and (\ref{phierr_eqn}) below).

In addition to the spectroscopic incompleteness, there is an additional surface brightness selection that is imposed on the SDSS main galaxy sample. As a part of their study of the luminosity function of low luminosity galaxies, \citet{blanton05} investigated the completeness as a function of surface brightness and found that the SDSS spectroscopic galaxy sample is greater than 90\% complete above a half light surface brightness of $\mu_{50,r}=22.4$ mag arcsec$^{-2}$ with the completeness dropping to 50\% at $\mu_{50,r}=23.4$ mag arcsec$^{-2}$. Since luminosity and surface brightness are correlated, the surface brightness selection preferentially selects against dwarf galaxies. For galaxies brighter than $M_r=-18$, \citet{blanton05} have a fit a gaussian to the surface brightness distribution in a series of absolute magnitude bins and have used this model to extrapolate the number of galaxies likely missed due to the surface brightness selection at fainter luminosities. The fraction of galaxies missing from the sample increases from near zero at $M_r=-18$ to approximately 40\% at $M_r=-16$.  However, these low luminosity, low surface brightness galaxies do not make a significant contribution to the total luminosity density. Even after correcting for the surface brightness incompleteness, about 90\% of the $r$-band luminosity density is due to galaxies with $M_r<-17$. Thus, the surface brightness selection is unlikely to affect our results for bright galaxies. However, we may be underestimating the number density of galaxies with $-18<M_r<-16$ by up to 40\%.

\subsection{Absolute magnitudes and volume densities}

We computed absolute magnitudes for our sample galaxies using, for example for the $r$-band,
\begin{equation}
M_{r,0.1} = m_r - 5 \log{D_L} - 25 - K_{0.1,r}(z) + (z - 0.1)Q
\label{absmag_eqn}
\end{equation} 
where $M_{r,0.1}$ is the absolute magnitude, $m_r$ is the extinction corrected $r$-band magnitude, $D_L$ is the luminosity distance in Mpc, $K_{0.1,r}(z)$ is the K-correction needed to account for the shifting of the galaxy SEDs with respect to the filter bandpass, and $Q$ is a term to account for luminosity evolution in units of magnitudes per redshift. A positive value for $Q$ means that galaxies get brighter with increasing redshift. Similar equations were used for the other bands. For calculating the luminosity distance we assumed a Hubble constant $H_0 = 70$ km s$^{-1}$ Mpc$^{-1}$ and a flat universe with matter density relative to the critical density of $\Omega_m=0.3$ and dark energy density of $\Omega_{\Lambda} = 0.7$. We calculated the K-corrections using the K\_CORRECT program, version 4.1.4, originally developed by \citet{blanton03a} and now extended to handle {\it GALEX} data \citep{blanton_roweis07}. Given the redshift of a galaxy, the K\_CORRECT program finds the linear combination of a set of template galaxy spectra that best reproduces the observed colors of a galaxy. The templates have already been determined as described by \citet{blanton_roweis07}. The coefficients in the linear combination for each galaxy were used later when determining the maximum volume out to which a given galaxy could be observed and be detected in the sample. As suggested by \citet{blanton03b,blanton03c}, we minimized the errors due to the K-corrections by correcting all of the galaxies to bandpasses shifted by $z=0.1$, a redshift near the median value for our sample. We denote absolute magnitudes and colors in this system with the subscript 0.1. 

Probably the most uncertain term in equation (\ref{absmag_eqn}) is the evolution term $Q$. When calculating the optical galaxy luminosity functions from SDSS data, \citet{blanton03b} fit for both luminosity and density evolution even within the relatively small redshift range sampled by the SDSS data. While the results for number and density evolution are highly correlated, the best fit luminosity functions are consistent with no number density evolution but significant luminosity evolution. In particular for the $r_{0.1}$ band, \citet{blanton03b} found $Q = 1.62 \pm 0.3$, corresponding to a difference of 0.4 mag over the redshift range in our sample of $0.01<z<0.25$. As shown by \citet{blanton03b}, neglecting evolution can lead to significant distortions in the shape of the luminosity function. We have adopted $Q=1.6$ for all bands considered here. This means that we have implicitly assumed that galaxies evolve in luminosity but not in color or in number density.

A value for $Q$ of 1.6 is roughly consistent with other determinations in the UV as well. \citet{schiminovich05} investigated the evolution of the UV luminosity density $\rho_{FUV}$ with redshift and found for $z<1$ that $\rho_{FUV} \propto (1+z)^{2.5\pm0.7}$, corresponding to $Q=1.9\pm0.5$, assuming that the increase in $\rho_{FUV}$ is due entirely to luminosity evolution. Based upon a sample of {\it GALEX} galaxies matched with redshifts from the Two Degree Field Redshift Survey \citep{colless01}, \citet{treyer05} measured evolution of the luminosity function characteristic magnitude $M^*$ of $\Delta M^*_{NUV} \approx 0.3 \pm 0.1$ mag between $z=0.05$ and $z=0.15$, equivalent to $Q=3\pm1$. These other measurements would tend to favor a somewhat larger evolution in the $UV$ with redshift as compared to the $r$-band. Indeed, galaxies most likely become bluer with increasing redshift as a result of the average star formation rates of galaxies increasing with redshift. For example, comparing a low-z sample from the SDSS and a sample of galaxies at $z\sim1$, \citet{blanton06} found that the blue sequence becomes bluer by 0.3 mag in $(u-r)$ while the red sequence becomes bluer by only 0.1 mag. In another study of the evolution of the blue and red sequences, \citet{faber06} analyzed the evolution of galaxies in the $(U-B)$ vs. $M_B$ CMD out to $z\sim1$. They found that $M^*_B$ becomes brighter by $\sim1.3$ mag out $z\sim1$ for both blue and red sequence galaxies. On the other hand, the luminosity function normalization $\phi^*$ for blue galaxies was found to be roughly constant while that of red galaxies is increasing. 

In order to asses the effect of color evolution on our results, we have recomputed the absolute magnitudes and volume densities described below for the $NUV$ sample assuming evolution in the $r$-band of $Q_r = 1.6$ and in the $NUV$ of $Q_{NUV}=3$. These choices correspond to a decrease in the $(NUV-r)$ color of a galaxy across our redshift range from $0.01 < z < 0.25$ of  0.34 mag. We have compared many of the results presented in the following sections with and without allowing for color evolution. The morphology of the color-magnitude diagram remains largely unchanged with the peaks and widths of the red and blue sequences nearly the same. The largest effect of including color evolution is on the volume density of luminous blue galaxies since these galaxies are detectable across the entire redshift range and thus would have the largest color correction. Including color evolution also tends to increase somewhat the volume density of galaxies with very blue colors $(NUV-r)<1$. Specifically, including color evolution would increase the luminosity density of the bluest galaxies with $0<(NUV-r)<1$ by 0.2 dex in the $r$-band and by 0.3 dex in the $NUV$. Due to the large uncertainties remaining in the evolution of galaxy colors with redshift and the relatively minor effect it has on our results, we decided to neglect color evolution in our analysis.  

When correcting for Galactic extinction, we assumed the \citet{cardelli89} extinction law with $R_V = A_V/E(B-V) = 3.1$. For the SDSS bands, the ratio of $A(\lambda)/E(B-V)$ is 5.155, 3.793,  and 2.751 for $u$, $g$, and $r$, respectively, while for the $FUV$ the ratio is 8.24. Due to the presence of the 2175~\AA~bump in the Galactic extinction law, the extinction in the $NUV$ band is no longer strictly proportional to the reddening $E(B-V)$. In order to quantify the effect this has on our extinction corrections, we used a small set of 42 SEDs from \citet{bruzual03} that span a representative range of galaxy SEDs from quiescent ellipticals to rapidly star-forming galaxies.\footnote{ The 42 SEDs can be found at http://www.lam.oamp.fr/arnouts/LE\_PHARE.html.} For each intrinsic SED, we applied the \citet{cardelli89} extinction law and then computed the resulting $NUV$ AB magnitude as a function of $E(B-V)$. For each SED, we fit a a quadratic function of $E(B-V)$:
\begin{equation}
A_{NUV} = a_1 E(B-V) + a_2 E(B-V)^2.
\end{equation}
For galaxies with some recent star formation, $a_1 = 8.24$ and $a_2=-0.67$ while for older galaxies with little or no recent star formation and a SED that falls steeply in the UV with decreasing wavelength, $a_1$ is slightly smaller and lies in the range $7.5-8.0$. For 97\% of our sample $E(B-V) < 0.1,$ and thus the quadratic term can be safely neglected. In addition, the maximum difference in the adopted value for $A(NUV)$ for the range of values for $a_1$ among the 42 SEDs, is only 0.07 mag at $E(B-V)=0.1$. Therefore, we assume the value $A(NUV)/E(B-V)=8.2$ for all of our calculations.

\begin{figure}
\includegraphics*[width=5in,height=5in]{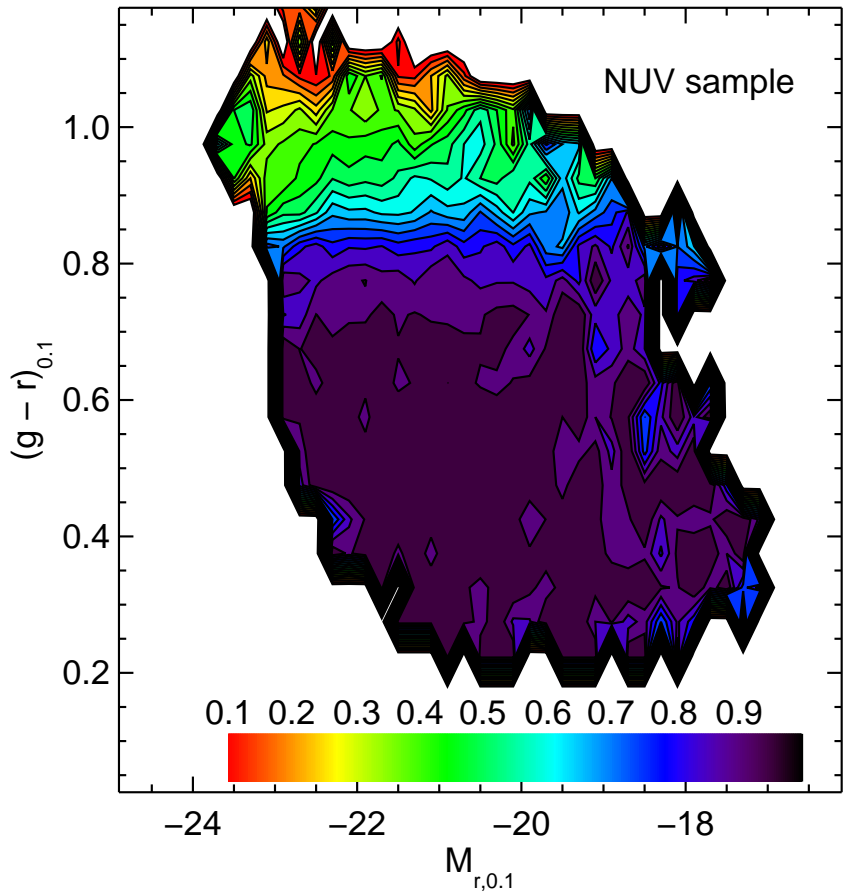}
\caption{The fraction of galaxies in the SDSS main galaxy sample with {\it GALEX} $NUV$ matches as a function of $(g-r)_{0.1}$ and $M_{r,0.1}$. The completeness for optical blue sequence galaxies is greater than $90\%$ while the completeness then drops for galaxies with colors $(g-r)_{0.1}>0.8$. For galaxies along the red sequence, the completeness in the $NUV$ is between $30-50$\%. This incompleteness is due to the UV magnitude limit and is accounted for when calculating the volume densities shown in the figures below.\label{completeness_nuv_gr}}
\end{figure}

\begin{figure}
\includegraphics*[width=5in,height=5in]{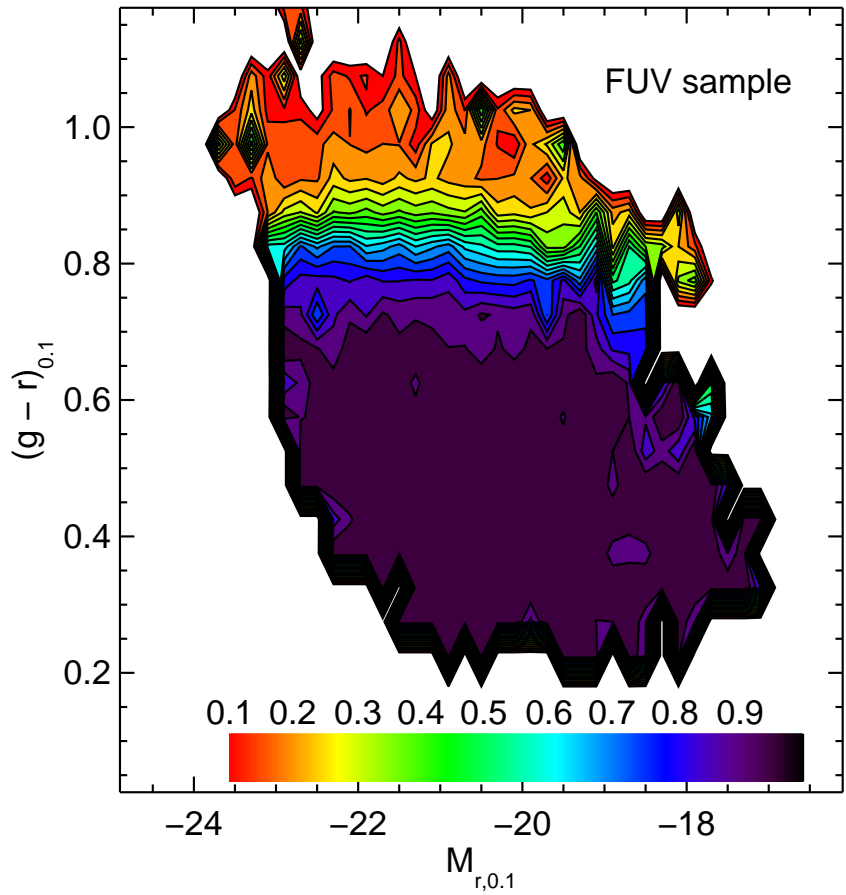}
\caption{Similar to Figure \ref{completeness_nuv_gr} except for the $FUV$ sample. The fraction of SDSS main galaxy sample galaxies with {\it GALEX} $FUV$ matches is about 90\% for blue galaxies and drops to values of $10-30$\% along the red sequence.\label{completeness_fuv_gr}}
\end{figure}

As argued above, the fraction of SDSS main sample galaxies with {\it GALEX} detections is a strong function of the galaxy color. In Figures \ref{completeness_nuv_gr} and \ref{completeness_fuv_gr}, we plot contours of the fraction of SDSS galaxies with a {\it GALEX} match for the $NUV$ and $FUV$ samples, respectively. In both samples, the completeness for blue galaxies is more than 90\% while the completeness begins to drop for galaxies with $(g-r)_{0.1}>0.8$. For the $NUV$ sample the completeness along most of the red sequence is in the range $30-60\%$ while the completeness of the red sequence in the $FUV$ sample is lower and in the range $10-40\%$. It is important to note that this drop-off in the fraction of galaxies with a {\it GALEX} match is due to the {\it GALEX} magnitude limit and was taken into account below when computing the volume densities of galaxies as a function of absolute magnitude and color.

We used the $V_{max}$ method \citep{schmidt68} to determine the volume densities of galaxies in our samples. The value of $V_{max}$ for each galaxy is given by the maximum volume within which the galaxy could have been included in the sample, given our selection limits listed in Table \ref{selection_limits}. We computed a separate $V_{max}$ for the $FUV$ and $NUV$ samples. First, we computed $K_{0.1}(z)$ for $0.01 <  z < 0.25$ for each galaxy using the best-fit SED derived from the output of the K\_CORRECT program. Next, we used equation (\ref{absmag_eqn}) to define a maximum and minimum redshift for each galaxy in each band by replacing the apparent magnitude in equation (\ref{absmag_eqn}) by the magnitude limits listed in Table \ref{selection_limits}. Then we computed a combined minimum and maximum redshift using
\begin{mathletters}
\begin{eqnarray}
z_{max} = min( z_{r,max}, z_{UV,max}, 0.25), \\
z_{min} = max(z_{r,min}, z_{UV,min}, 0.01).
\end{eqnarray}
\end{mathletters}
Since we have assumed a cosmology with no overall curvature, we calculated the volume between $z_{min}$ and $z_{max}$ for each galaxy as
\begin{equation}
V_{max} = \frac{A}{3} \left(\frac{\pi}{180}\right)^2 \left( \frac{D_L(z_{max})^3}{(1+z_{max})^3} - \frac{D_L(z_{min})^3}{(1+z_{zmin})^3}\right),
\end{equation}
where $A$ is the solid angle in deg$^2$ from which the sample was drawn. The $V_{max}$ values can then be used to generate the number densities of galaxies as a function of any variables. For example, in order to generate a volume-corrected galaxy CMD for the $NUV$ sample, we computed the number density of galaxies as a function of absolute magnitude and color using 
\begin{equation}
\Phi(M_{r,0.1},(NUV-r)_{0.1}) = \frac{f}{\Delta M \Delta C} \sum \frac{1}{V_{max}},
\label{phi_eqn}
\end{equation}
where $\Phi$ gives the number density of galaxies in units of Mpc$^{-3}$ mag$^{-2}$, $\Delta M$ is the width of each bin in absolute magnitude, $\Delta C$ is the width of each bin in color, $f$ is the inverse of the sample completeness estimated above in \S2.2, and the sum is taken over all galaxies within that particular color and absolute magnitude bin centered on $M_{r,0.1}$ and $(NUV-r)_{0.1}$. The corresponding uncertainty in each bin due to counting statistics is given by
\begin{equation}
\delta \Phi(M_{r,0.1},(NUV-r)_{0.1}) = \frac{f}{\Delta M \Delta C} \left( \sum \frac{1}{V_{max}^2}\right)^{1/2}.
\label{phierr_eqn}
\end{equation}

\section{Results}

\subsection{The Galaxy Color Magnitude Diagram}

\begin{figure}
\includegraphics*[width=5in,height=5in]{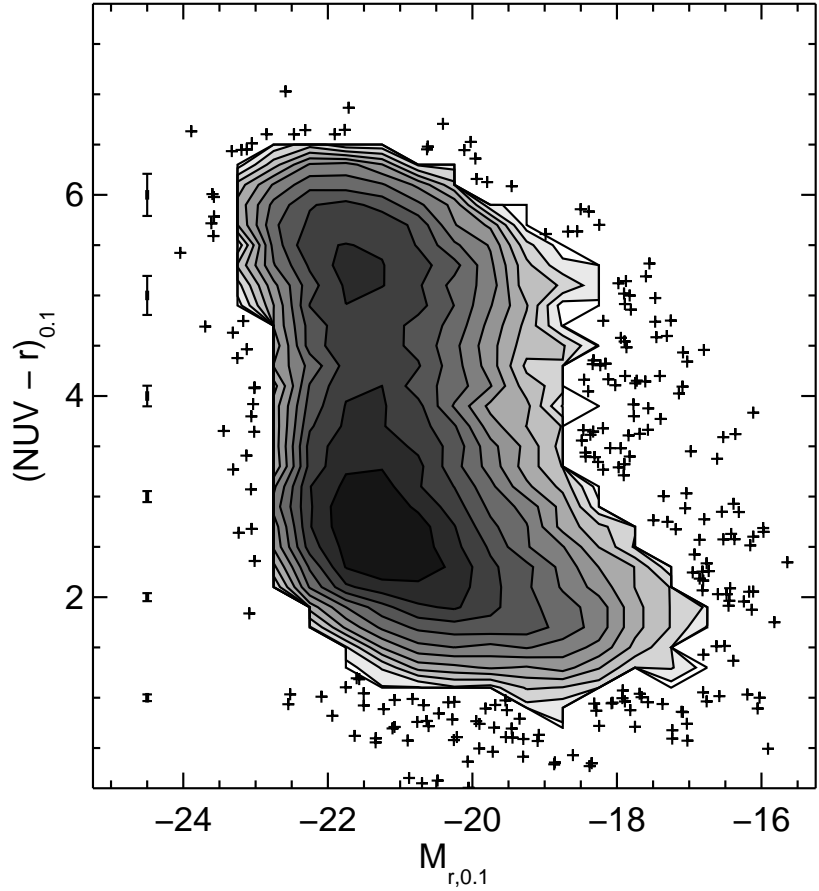}
\caption{The distribution of galaxies in the $(NUV - r)_{0.1}$ vs. $M_{r,0.1}$ plane. The contours are spaced logarithmically from a minimum of five galaxies per bin up to 300 per bin. Below the minimum contour level, the positions of individual galaxies are plotted. The median photometric errors as a function of color are shown along the left-hand side of the plot.\label{cmd_nuv_points}}
\end{figure}

\begin{figure}
\includegraphics*[width=5in,height=5in]{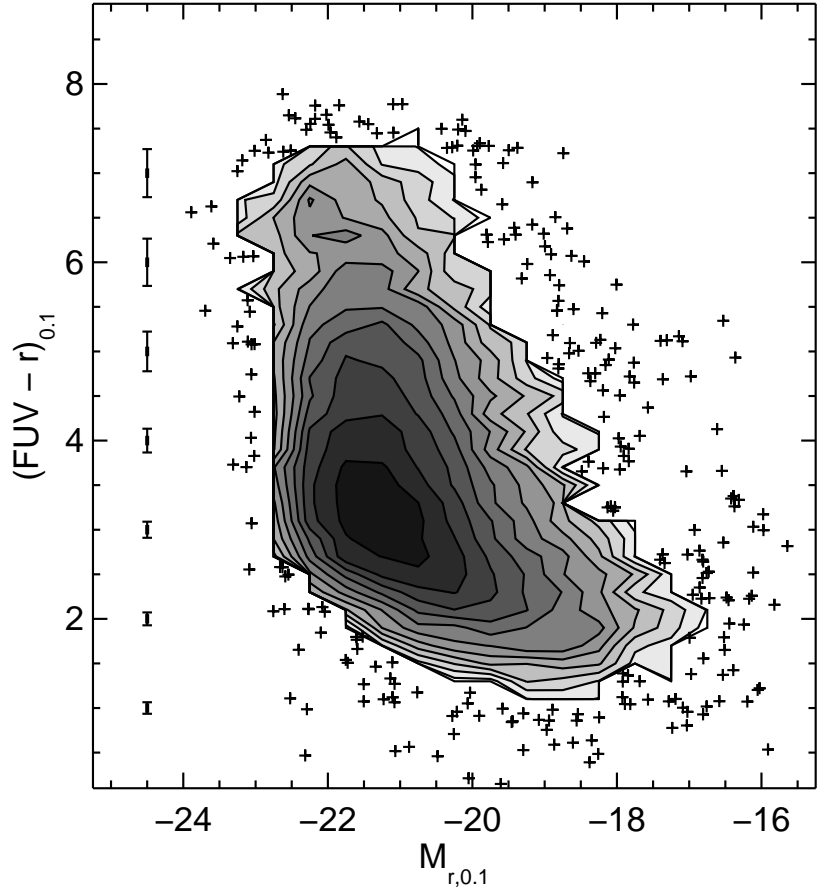}
\caption{The same as Figure \ref{cmd_nuv_points} except for $(FUV - r)_{0.1}$ vs. $M_{r,0.1}$. The data are plotted in bins that are 0.2 mag wide in color and 0.5 mag wide in absolute magnitude. The contours are spaced logarithmically from a minimum of five galaxies per bin up to a maximum of 250 per bin. The median photometric errors as a function of color are shown along the left hand side of the plot. \label{cmd_fuv_points}}
\end{figure}

The number of galaxies in our $(NUV-r)_{0.1}$ and $(FUV-r)_{0.1}$ CMDs are plotted in Figures \ref{cmd_nuv_points} and \ref{cmd_fuv_points}, respectively. In both plots, the data are plotted as contours where the density of points is high while the locations of individual galaxies are plotted where the density is low.  The uncertainty in the colors in both samples is dominated by the errors in the UV measurements. Thus the errors as a function of position in the CMDs are most strongly correlated with color. The median error as a function of color is plotted in Figures \ref{cmd_nuv_points} and \ref{cmd_fuv_points} along the left-hand side of each figure. For blue galaxies, the uncertainty is dominated by the zero-point uncertainty in the {\it GALEX} data. For red galaxies, the errors span a larger range from $0.1-0.4$ mag with a median of about 0.2 mag. 

\begin{figure}
\includegraphics*[width=5in,height=5in]{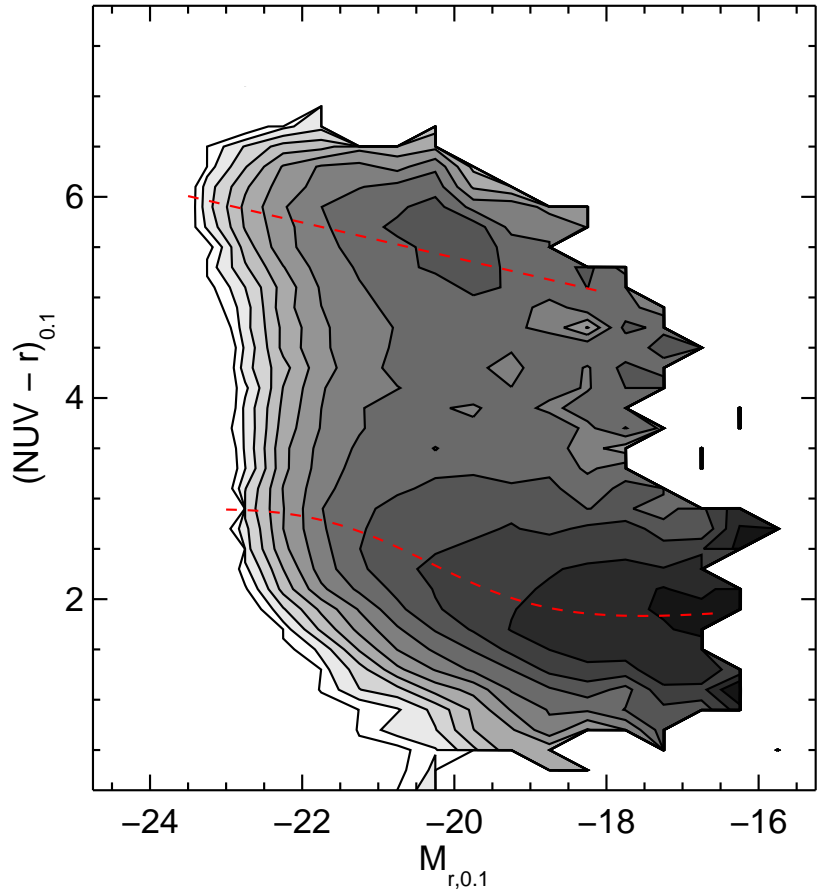}
\caption{The volume density of galaxies as a function of
$(NUV-r)_{0.1}$ and $M_{r,0.1}$ using the $V_{max}$ method. The
density is determined in 0.2 mag wide bins in color and 0.5 mag wide
bins in absolute magnitude. There are twelve logarithmically spaced contours ranging
from $10^{-5.5}$ to $10^{-2.3}$ Mpc$^{-3}$ mag$^{-2}$.
The dashed lines indicate the ridge lines of the red and blue sequences as determined from the Gaussian fits shown in Figure \ref{cmd_nuv_seq} below.
\label{cmd_nuv_vmax}}
\end{figure}

\begin{figure}
\includegraphics*[width=5in,height=5in]{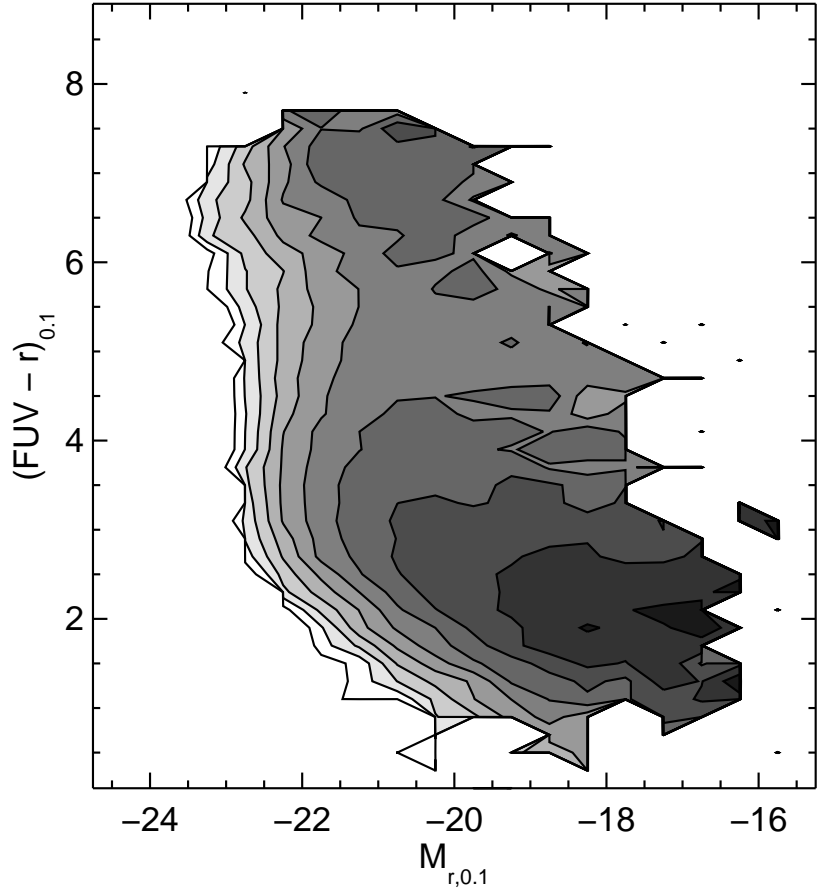}
\caption{The volume density of galaxies as a function of
$(FUV-r)_{0.1}$ and $M_{r,0.1}$ using the $V_{max}$ method. The
density is determined in 0.2 mag wide bins in color and 0.5 mag wide
bins in absolute magnitude. The contours are logarithmically spaced
from $10^{-5.5}$ to $10^{-2.3}$ Mpc$^{-3}$ mag$^{-2}$. 
\label{cmd_fuv_vmax}}
\end{figure}

The corresponding volume densities of galaxies as a function of position in the CMD are plotted in Figures \ref{cmd_nuv_vmax} and \ref{cmd_fuv_vmax}, where the weighting was derived from the $V_{max}$ values as in equation (\ref{phi_eqn}) with $\Delta M = 0.5$ mag and $\Delta C = 0.2$ mag. The peak of each sequence from the Gaussian fits described below are over-plotted as the dashed lines in the $(NUV-r)$ diagram. The volume densities and the errors for the $(NUV-r)$ diagram are given in Tables \ref{cmd_nuv_tab} and \ref{cmderr_nuv_tab} while those for the $(FUV-r)$ diagram are given in Tables \ref{cmd_fuv_tab} and \ref{cmderr_fuv_tab}.

As we have argued in \S2.2, the red edge of the color distribution in the $NUV$ sample is real whereas the red edge of the $FUV$ sample is a selection effect due to the $FUV$ flux limit. This is reflected in the morphology of the galaxy distributions in Figures \ref{cmd_nuv_vmax} and \ref{cmd_fuv_vmax}, where the distribution turns over for the reddest colors in the $NUV$ diagram and does not turn over entirely in the $FUV$ diagram. Throughout the remainder of this paper, we focus on the $NUV$ diagram.

In both the $FUV$ and $NUV$ diagrams, the galaxies separate into two well-defined sequences in addition to a population of galaxies that lie in between. As has been noted before in optical CMDs \citep[e.g.][]{baldry04}, the most luminous galaxies are on the red sequence, while both sequences become redder with increasing luminosity. In contrast to the optical $(u-r)$ CMD, the blue sequence does not appear to merge at the bright end with the red sequence.

\begin{figure}
\includegraphics*[width=5in,height=5in]{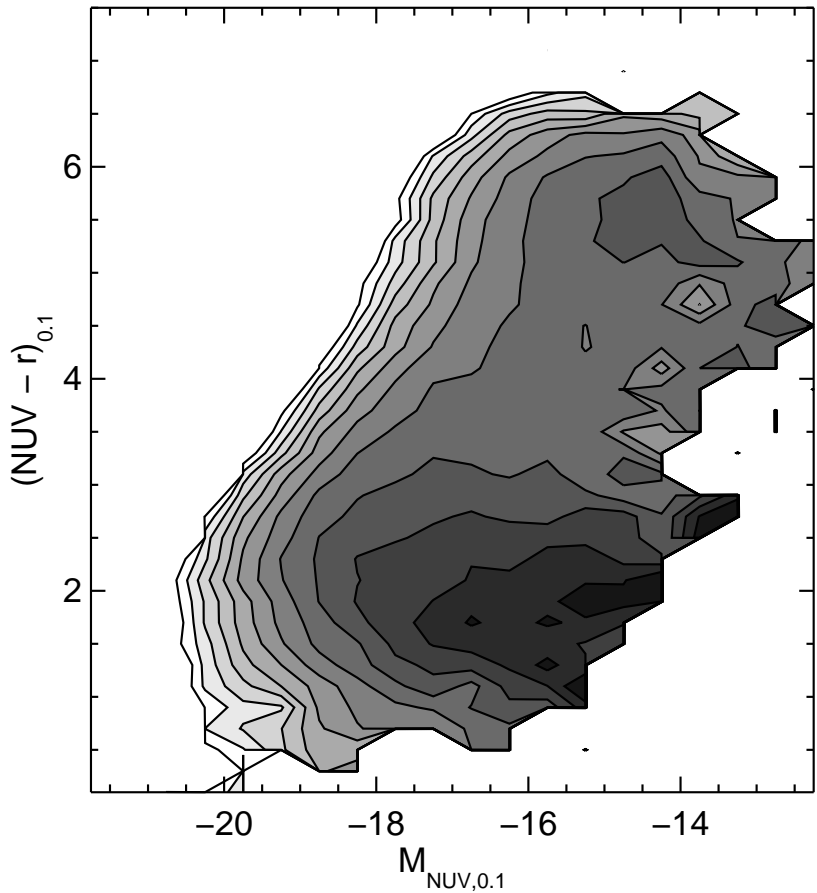}
\caption{The volume density of galaxies as a function of $(NUV-r)_{0.1}$ and $M_{NUV,0.1}$ using the $V_{max}$ method. The density is determined in 0.2 mag wide bins in color and 0.5 mag wide bins in absolute magnitude. The contours are logarithmically spaced from  $10^{-5.5}$ to $10^{-2.3}$ Mpc$^{-3}$ mag$^{-2}$. This is the same sample as in Figure \ref{cmd_nuv_vmax} except plotted as a function of the $NUV$ absolute magnitude. The sample reaches significantly fainter $NUV$ absolute magnitudes for the red galaxies due to the SDSS $r$-band magnitude limit. \label{cmd_nuv_vmax2}}
\end{figure}

An alternative view of the $NUV$ sample is shown in Figure \ref{cmd_nuv_vmax2} where the volume density of galaxies is plotted as a function of the $NUV$ luminosity. The sample reaches significantly fainter $NUV$ absolute magnitudes for the red galaxies due to the SDSS $r$-band selection. Thus, the slope for the faintest $NUV$ absolute magnitudes included in our sample as a function of color is a selection effect. As in Figure \ref{cmd_nuv_vmax}, the sample separates into blue and red sequences. However, there is little, if any, trend of color with $M_{NUV,0.1}$ along either sequence.  This is consistent with the conclusions from studies in the optical which indicate that one of the most important factors determining the evolution of a galaxy is its mass, which is much more closely related to the $r$-band luminosity than to the $NUV$ luminosity \citep{kauffmann03b,brinchmann04}.

\subsection{Color Distributions as a Function of $M_{r,0.1}$}
 
\begin{figure}
\includegraphics*[height=5in,keepaspectratio=1]{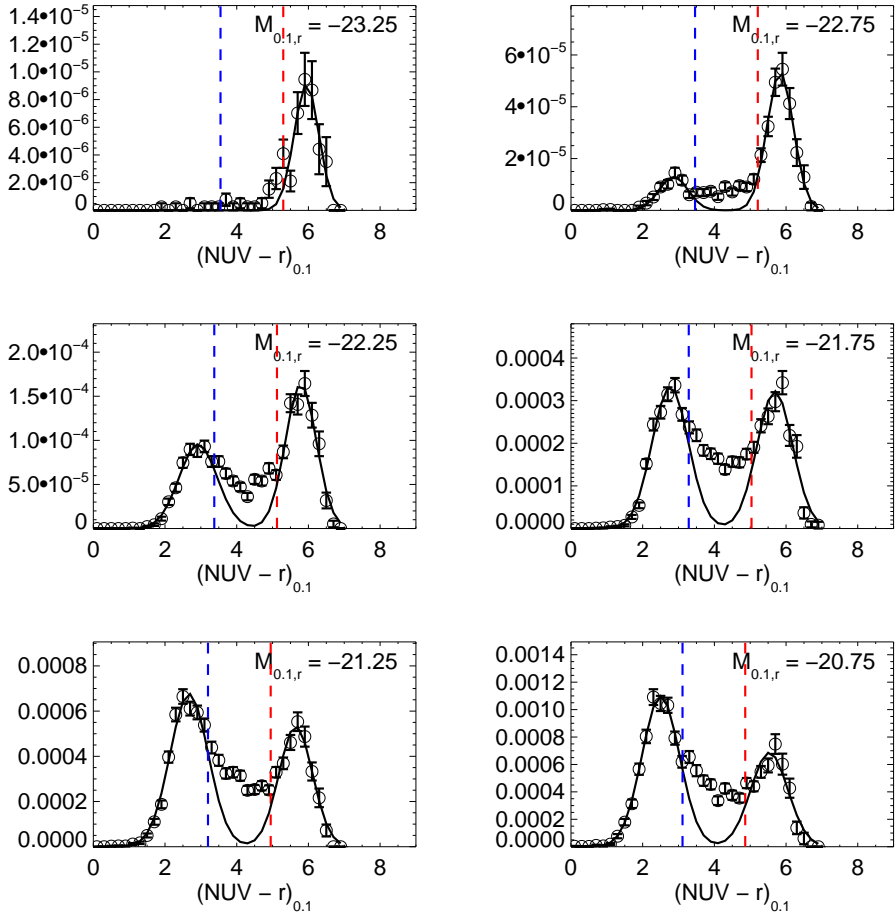}
\caption{The volume density of galaxies from the $V_{max}$ method as a function of $(NUV -
r)_{0.1}$ color in 0.5 magnitude wide absolute magnitude bins. The
midpoint of each absolute magnitude bin is labeled in the upper right
of each figure.  The blue and red dashed lines mark the limits used when 
fitting Gaussians to the red and blue sequences separately. The solid black
line is the sum of the two Gaussians. Note that in most absolute magnitude
bins, the color distribution is not well fit by the sum of two
Gaussians as there is a significant excess of objects in between the
two sequences.\label{cmd_nuv_colordist}}
\end{figure}

\begin{figure}
\includegraphics*[height=5in,keepaspectratio=1]{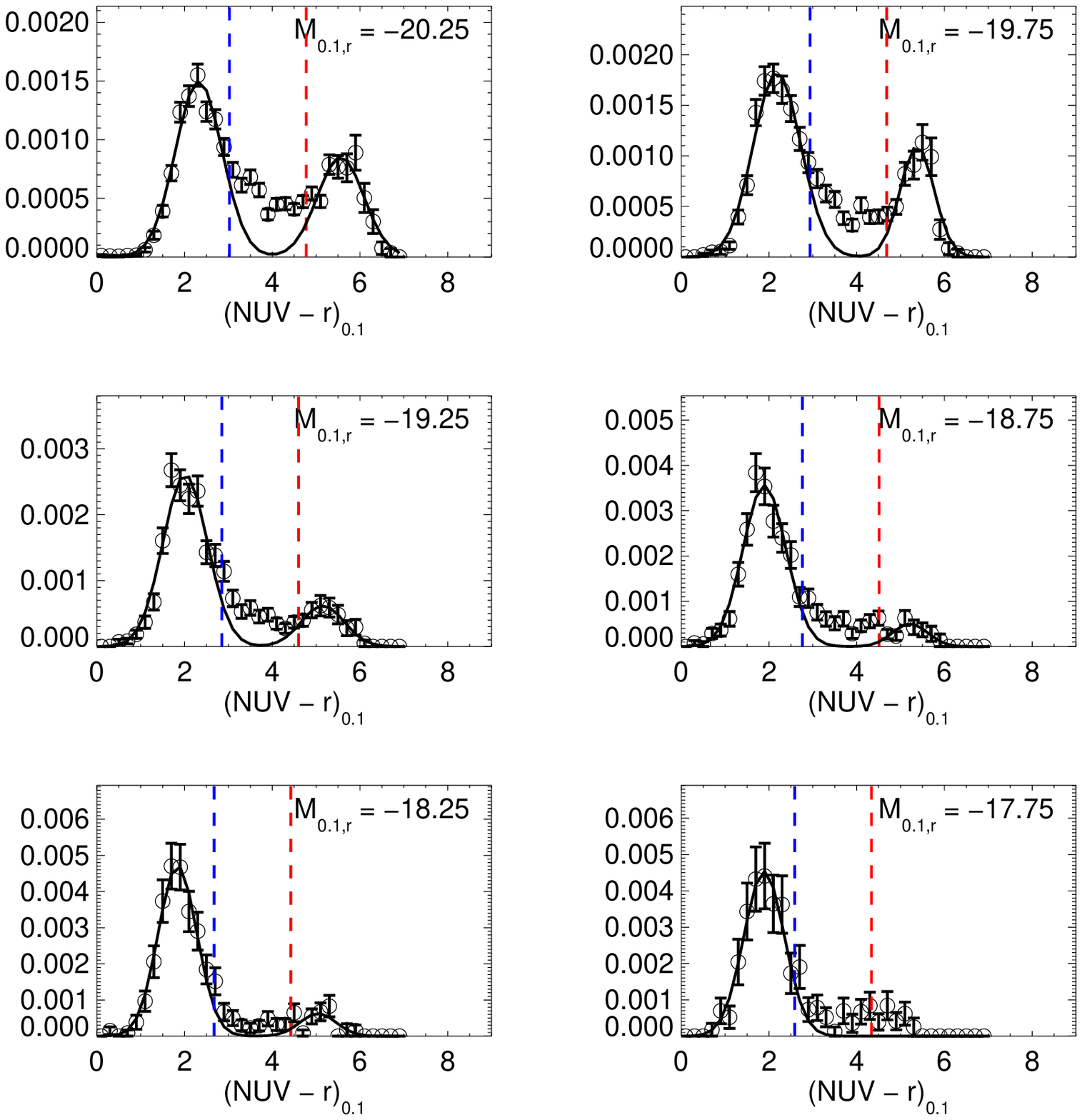}
\caption{The continuation of the previous figure.\label{cmd_nuv_colordist2}}
\end{figure}

\begin{figure}
\includegraphics*[height=5in,keepaspectratio=1]{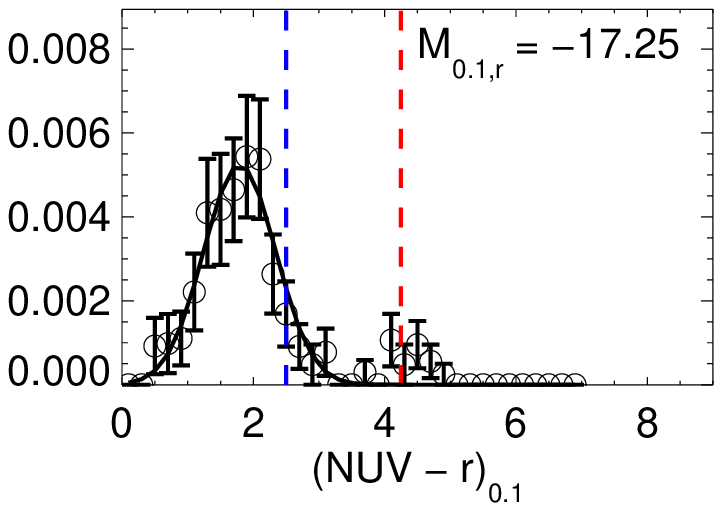}
\caption{The continuation of the previous figure.\label{cmd_nuv_colordist3}}
\end{figure}

The volume-corrected number density of galaxies as a function of $(NUV-r)_{0.1}$ color is plotted in Figures \ref{cmd_nuv_colordist}$-$\ref{cmd_nuv_colordist3} in 0.5 magnitude wide bins of $M_{r,0.1}$. The error bars are the statistical errors only calculated using equation (\ref{phierr_eqn}). Except for the most luminous bin, there is both a red and blue peak visible in each panel. Similar to previous optical CMDs, the red sequence dominates in the brighter bins. The red and blue sequences reach approximately equal strengths around $M_{r,0.1}=-21.75$ with the blue sequence becoming dominant at fainter luminosities. The relative number of red sequence galaxies reaches 50\% at about the same luminosity when dividing galaxies using the $(u-r)$ color \citep{baldry04}.
 
Following \citet{baldry04}, we attempted to fit Gaussians to the red and blue peaks in the color distributions in each $M_{r,0.1}$ bin although we employed a somewhat different methodology. We fit a single Gaussian of the form $(1/\sqrt{2\pi\sigma^2}) \exp{\{-((NUV-r)_{0.1}-\mu)^2/2\sigma^2\}}$ separately to the red and blue sequences. In order to select points on the red and blue sequences, we have utilized the fit from \citet{yi05} to the $(NUV-r)$ color as a function of $M_r$ for a sample of morphologically selected early-type galaxies: $(NUV-r) = f(M_r) = 1.73 - 0.17 M_r$. For the purposes of fitting a Gaussian to each sequence, we defined the red sequence as the points with $(NUV-r)_{0.1} > f(M_{r,0.1}) - 0.5$ and fit a Gaussian to these points using the IDL routine GAUSSFIT, a program that computes a non-linear least squares fit to the data. Similarly, we fit a Gaussian to the blue sequence for points with $(NUV-r)_{0.1} < f(M_{r,0.1}) - 2.0$. These color limits are plotted as the dashed red and blue lines in each of the color distributions shown in Figures \ref{cmd_nuv_colordist}$-$\ref{cmd_nuv_colordist3} whereas the sum of the best-fitting Gaussians is plotted as the solid line. In contrast to the optical $(u-r)$ CMD, a double-Gaussian does not provide a good fit to the data. Although the Gaussians provide a reasonable fit to the blue edge of the blue sequence and the red edge of the red sequence, their sum falls well below the data in the region between the two sequences. 

For the star-forming galaxies in the blue sequence, it is difficult to generate galaxies with extremely blue colors, i.e. $(NUV-r)_{0.1} \lesssim 1$, except with large starbursts or very young ages \citep[e.g.][]{treyer98}. Clearly, such objects are rare in the local universe. The skew of the blue sequence in the $(NUV-r)_{0.1}$ CMD to redder colors compared to what is observed in the optical would be expected for galaxies with somewhat older average stellar populations or with larger reddening due to dust. Similarly, whereas the red sequence is relatively narrow in the optical, a color distribution skewed towards the blue would be expected for early-type galaxies with some residual star formation \citep{yi05}. The departure of the blue and red sequences from a Gaussian would imply that we are beginning to resolve some of these effects due to the greater sensitivity of the UV light to both variations in the star formation rate and the amount of dust. 
 
\begin{figure}
\includegraphics*[width=5in,height=5in]{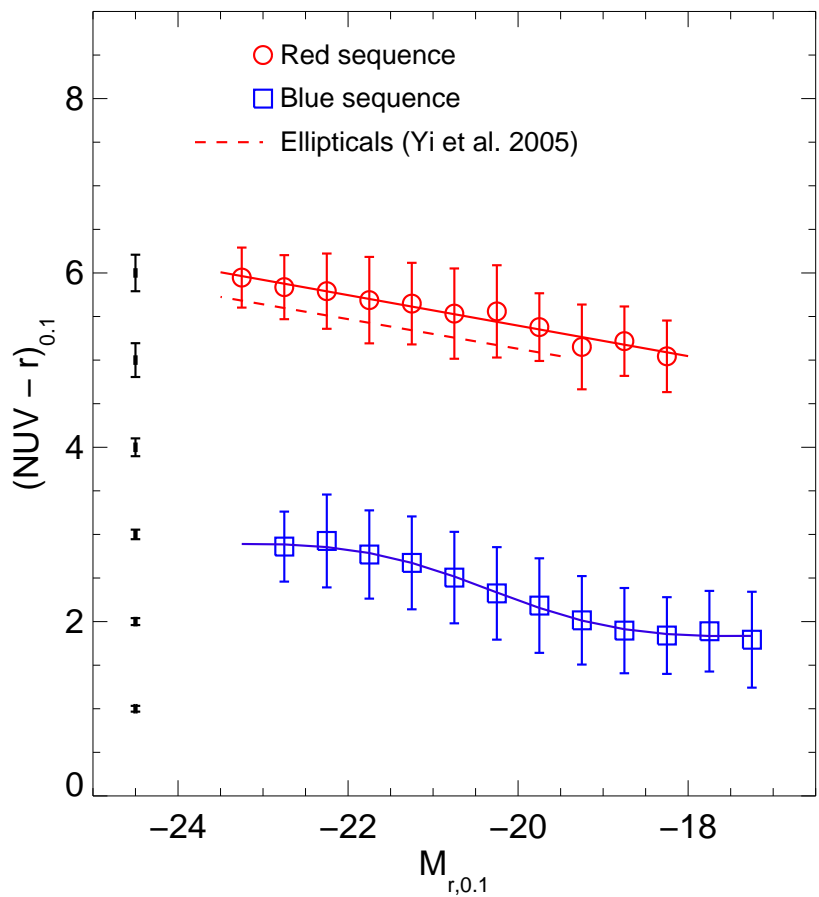}
\caption{Plot of the Gaussian fits to the red and blue sequences in
Figures \ref{cmd_nuv_colordist}$-$\ref{cmd_nuv_colordist3}. The red circles and blue squares refer
to the means of the Gaussians for the red and blue sequences,
respectively, while the error bars on each point refer to the width
$\sigma$ for the fit to each sequence. The solid red line is a simple
linear fit to the red sequence while the dashed red line indicates the fit to the
color as a function of absolute magnitude from \citet{yi05} for
morphologically selected elliptical galaxies. The solid
blue line is the best-fit $T$-function to the blue sequence. The median
photometric errors as a function of color are shown along the left hand side
of the plot.
\label{cmd_nuv_seq}}
\end{figure}

Even though a double Gaussian provides a poor fit to the $(NUV-r)_{0.1}$ color distribution, the peak of each Gaussian provides a robust estimate of the peak of each sequence in each absolute magnitude bin. On the other hand, the width $\sigma$ of each Gaussian should be interpreted with caution as it only gives some information about the blue edge of the blue sequence and the red edge of the red sequence and does not provide a good representation of the entire distribution.  The resulting parameters of the Gaussian fits are plotted in Figure \ref{cmd_nuv_seq}. In the figure the circles and squares give the peaks of the red and blue sequences, respectively, whereas the error bars denote the $\sigma$ for each Gaussian. The values of $\sigma$ for the red sequence lie in the range $0.3-0.5$ mag while those for the blue sequence lie in the range $0.5-0.6$ mag. Also plotted in Figure \ref{cmd_nuv_seq} are the median photometric errors as a function of color for comparison.  While the values for $\sigma$ for the blue sequence are significantly larger than the photometric errors, the values for the red sequence are comparable to the photometric errors in the color, indicating that the fall-off of the color distribution on the red edge of the red sequence is largely consistent with that expected from the errors.

We have fit a line to the peak color of the red sequence as a function of absolute magnitude with the result $(NUV-r)_{0.1} = 1.897 - 0.175 M_{r,0.1}$. This fit is plotted as the dashed red line in Figure \ref{cmd_nuv_vmax} and the solid red line in Figure \ref{cmd_nuv_seq}. This fit has the same slope as found by \citet{yi05} except with a slightly redder intercept. \citet{yi05} analyzed a sample of morphologically selected early type galaxies, some of which appear to harbor some residual star formation, a fact which would tend to pull their fit somewhat towards the blue compared to our color-selected sample. 

Following \citet{baldry04}, we fit the peak color of the blue sequence with the sum of a line plus a tanh function with the result
\begin{equation}
(NUV-r)_{0.1}  = 2.39 + 0.075 (M_{r,0.1} + 20) - 0.808 \tanh{\left( \frac{M_{r,0.1} + 20.32}{1.81}\right)}.
\end{equation}
This curve is overplotted as the solid blue line in Figure \ref{cmd_nuv_seq}. Overall the peak of the blue sequence increases in color from $(NUV-r)_{0.1}=1.8$ at the faint end to $\sim 3$ for the most luminous blue galaxies. The color of the blue sequence changes most rapidly around $M_{r,0.1}=-20.3$.
 
\subsection{Luminosity Functions as a Function of Color}

\subsubsection{$M_{r,0.1}$ luminosity functions}

\begin{figure}
\includegraphics*[width=5in,height=5in]{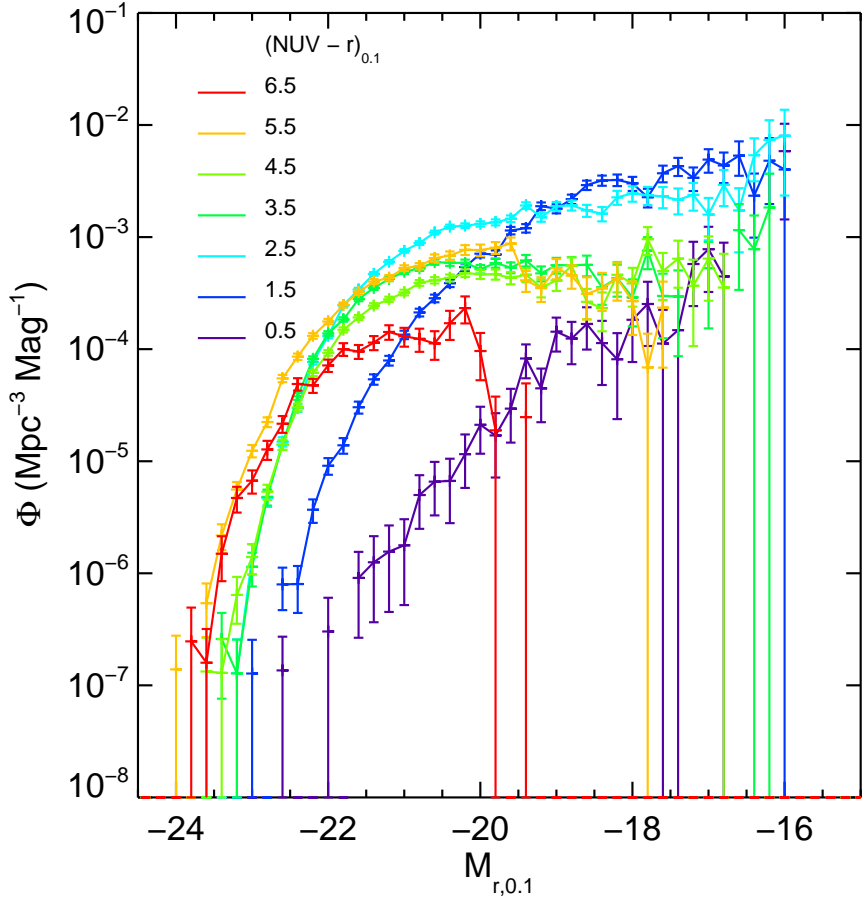}
\caption{The $M_{r,0.1}$ luminosity functions separated in one
magnitude wide bins of $(NUV - r)_{0.1}$ color. The errors bars
represent the statistical error derived from the $V_{max}$
calculation. \label{cmd_nuv_lfs}}
\end{figure}

\begin{figure}
\includegraphics*[width=5in,height=5in]{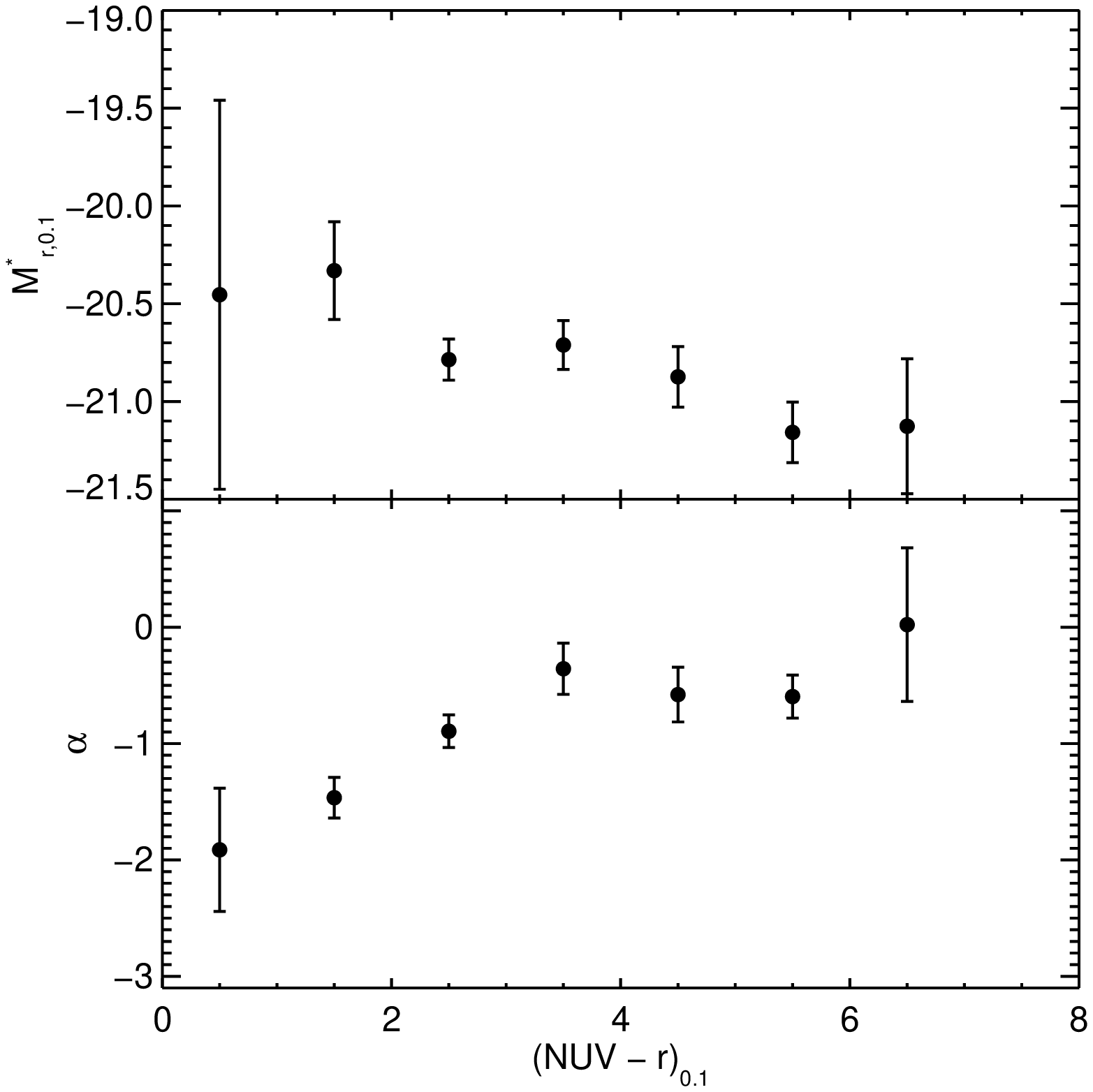}
\caption{The best-fit values for $M^{*}_{r,0.1}$ (top panel) and
$\alpha$ (bottom panel) as a function of $(NUV-r)_{0.1}$ color from
Schechter function fits to the luminosity functions plotted in Figure
\ref{cmd_nuv_lfs}. The error bars are derived from the range of
parameters within one of the minimum reduced $\chi^2$.
\label{cmd_nuv_lfs_param}}
\end{figure}

The galaxy luminosity function varies strongly with color even within each sequence. This is illustrated in Figure \ref{cmd_nuv_lfs} where the $M_{r,0.1}$ luminosity functions are plotted separately for one magnitude wide bins in $(NUV-r)_{0.1}$ color ranging from zero to seven. These are essentially horizontal cuts through the CMD in Figure \ref{cmd_nuv_vmax}. We have fit the luminosity function in each color bin with a \citet{schechter76} function given by
\begin{equation}
\Phi(M) = 0.4 \ln{10} \phi^* 10^{-0.4(M-M^*)(\alpha+1)} \exp{\{-10^{-0.4(M-M^*)}\}},
\label{schechter}
\end{equation}
where $\phi^*$, $M^*$, and $\alpha$ are fit to each sequence by minimizing $\chi^2$. We determined the $1\sigma$ errors in each of the parameters using the range of solutions within one of the minimum reduced $\chi^2$. The best-fit parameters are listed in Table \ref{schechter_r} while $M^*_{r,0.1}$ and $\alpha$ are plotted as a function of $(NUV-r)_{0.1}$ in Figure \ref{cmd_nuv_lfs_param}.

The luminosity function for the very bluest bin with $0<(NUV-r)_{0.1}<1$ is very steep with a best-fitting faint end slope of nearly $-2$, although the errors on $M^*$ and $\alpha$ for this bin are large and highly correlated. As the color increases through the blue sequence, the faint end slope $\alpha$ gradually increases, reaching values of $\sim -0.5$ in between the two sequences at $(NUV-r)_{0.1}\approx4$. Although the uncertainties become larger, the slope reaches a slighty larger value of $\sim0$ for the reddest bin. The value of $M^*$ similarly varies systematically with color, going from $-20.4$ in the bluest bins to $-20.8$ at intermediate colors and finally reaching $-21.1$ for the reddest galaxies. Qualitatively similar results were found by \citet{blanton01} when determining the $r$-band luminosity function separated by $(g-r)$ color. 

We have computed the total luminosity density within each color bin from the best-fit Schechter parameters as $\rho_{r,0.1}=\int^{\infty}_{0}L\Phi(L)dL = \phi^* L^* \Gamma(\alpha+2)$. The statistical errors in $\rho$ were determined using the range of values corresponding to those fits within one of the minimum reduced $\chi^2$. For values of the faint end slope $\alpha<-2$, the integral of the Schechter function diverges when integrating all the way down to zero luminosity. With the exception of the bluest bin, the luminosity function slopes are significantly larger than $-2$. However, the $1\sigma$ confidence interval for $\alpha$ for the bluest bin includes values in the region where $\alpha<-2$. When computing the luminosity density in this bin, we have integrated the luminosity function only down to $M_{r,0.1}=-12$. Since this choice is somewhat arbitrary, the luminosity density of the bluest galaxies is more uncertain than for the other color bins. However, we note that this luminosity reaches about the limits for which the galaxy luminosity function has been determined from the SDSS \citep{blanton05}.

\begin{figure}
\includegraphics*[width=5in,height=5in]{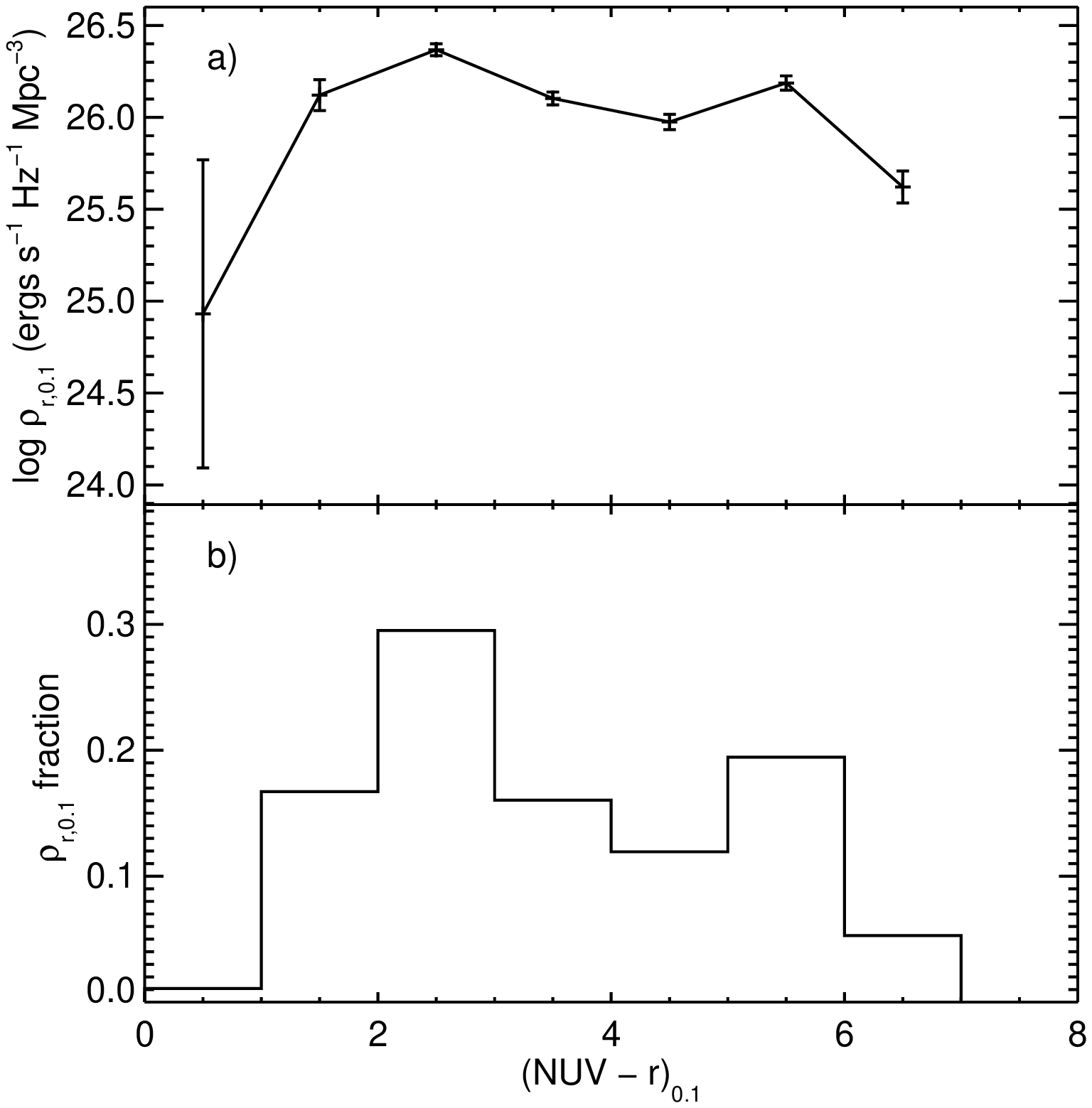}
\caption{(a) The $r_{0.1}$ band luminosity density $\rho_{r,0.1}$ as a function of $(NUV-r)_{0.1}$ color calculated by integrating the luminosity functions using the best-fit Schechter function parameters listed in Table \ref{schechter_r}. (b) The fraction of the total luminosity density contributed from each color bin. \label{lumden_r}}
\end{figure}

The luminosity densities are listed in Table \ref{schechter_r} and are plotted as a function of color in the top panel of Figure \ref{lumden_r} while the fraction of the total luminosity density within each color bin is plotted in the bottom panel of the figure. The largest contribution to the luminosity density comes from galaxies with $2<(NUV-r)_{0.1}<3$ and accounts for $\approx30\% $ of the total. Galaxies bluer than $(NUV-r)_{0.1}=4$ together contribute 64\% of the luminosity density while redder galaxies account for 36\%. Adding up the the contribution to the luminosity density from each color bin, we obtain a total of $\log{\rho_{r,0.1}}=26.903\pm0.030$ ergs s$^{-1}$ Hz$^{-1}$ Mpc$^{-3}$. This is only slightly larger than the luminosity density of $\log{\rho_{r,0.1}}=26.845\pm0.012$ calculated from a much larger sample of SDSS galaxies by \citet{blanton03b}, after converting to our value for the Hubble constant.

\subsubsection{$M_{NUV,0.1}$ luminosity functions}

\begin{figure}
\includegraphics*[width=5in,height=5in]{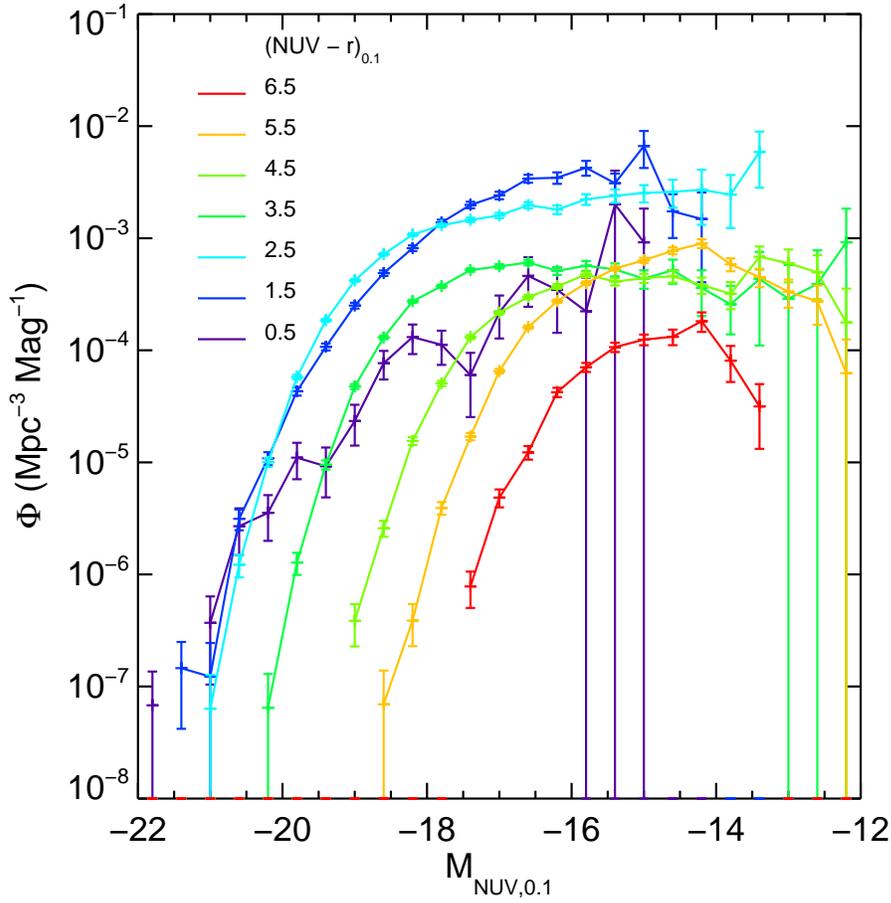}
\caption{The $M_{NUV,0.1}$ luminosity functions for galaxies in one magnitude wide bins in color derived using the $V_{max}$ method.\label{cmd_nuv_lfs2}}
\end{figure}

\begin{figure}
\includegraphics*[width=5in,height=5in]{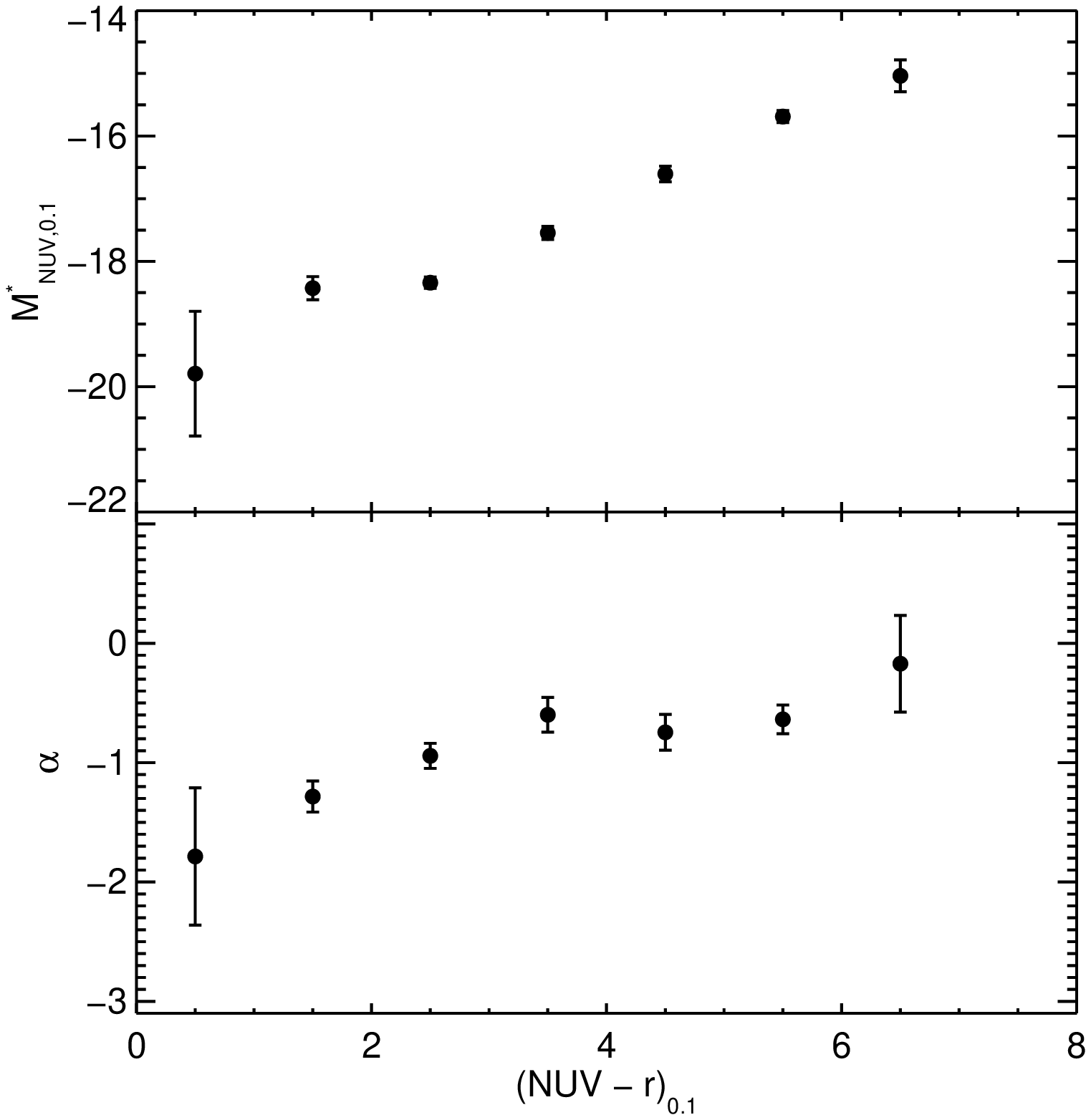}
\caption{The best-fit values for $M^{*}_{NUV,0.1}$ (top panel) and
$\alpha$ (bottom panel) as a function of $(NUV-r)_{0.1}$ color from
Schechter function fits to the luminosity functions plotted in Figure
\ref{cmd_nuv_lfs2}. The error bars are derived from the range of
parameters within one of the best-fit $\chi^2$.\label{cmd_nuv_lfs_param2}}
\end{figure}

Similar to the analysis of the $M_{r,0.1}$ luminosity functions described in the previous section, we have computed $M_{NUV,0.1}$ luminosity functions in one magnitude wide bins of $(NUV-r)_{0.1}$ color. As for the $r$-band, we fit Schechter functions to the distribution within each color bin. The luminosity functions are plotted in Figure \ref{cmd_nuv_lfs2} while the best-fitting Schechter function parameters are listed in Table \ref{schechter_nuv}. The best-fit values for $M^*_{NUV,0.1}$  and $\alpha$ are plotted as a function of color in Figure \ref{cmd_nuv_lfs_param2}. Qualitatively, the results are similar to that in the $r$-band. The faint end slope $\alpha$ is very steep for the bluest galaxies with a value of $-1.8$. The faint end slope gradually increases with color up to a value of $-0.6$ at $(NUV-r)_{0.1}=3.5$. For redder galaxies, the faint end slope remains at about this value until increasing slightly again in the reddest bin. The value of $M^*_{NUV,0.1}$ increases dramatically from $\sim-20$ for the bluest galaxies to $\sim-15$ for the reddest.

\begin{figure}
\includegraphics*[width=5in,height=5in]{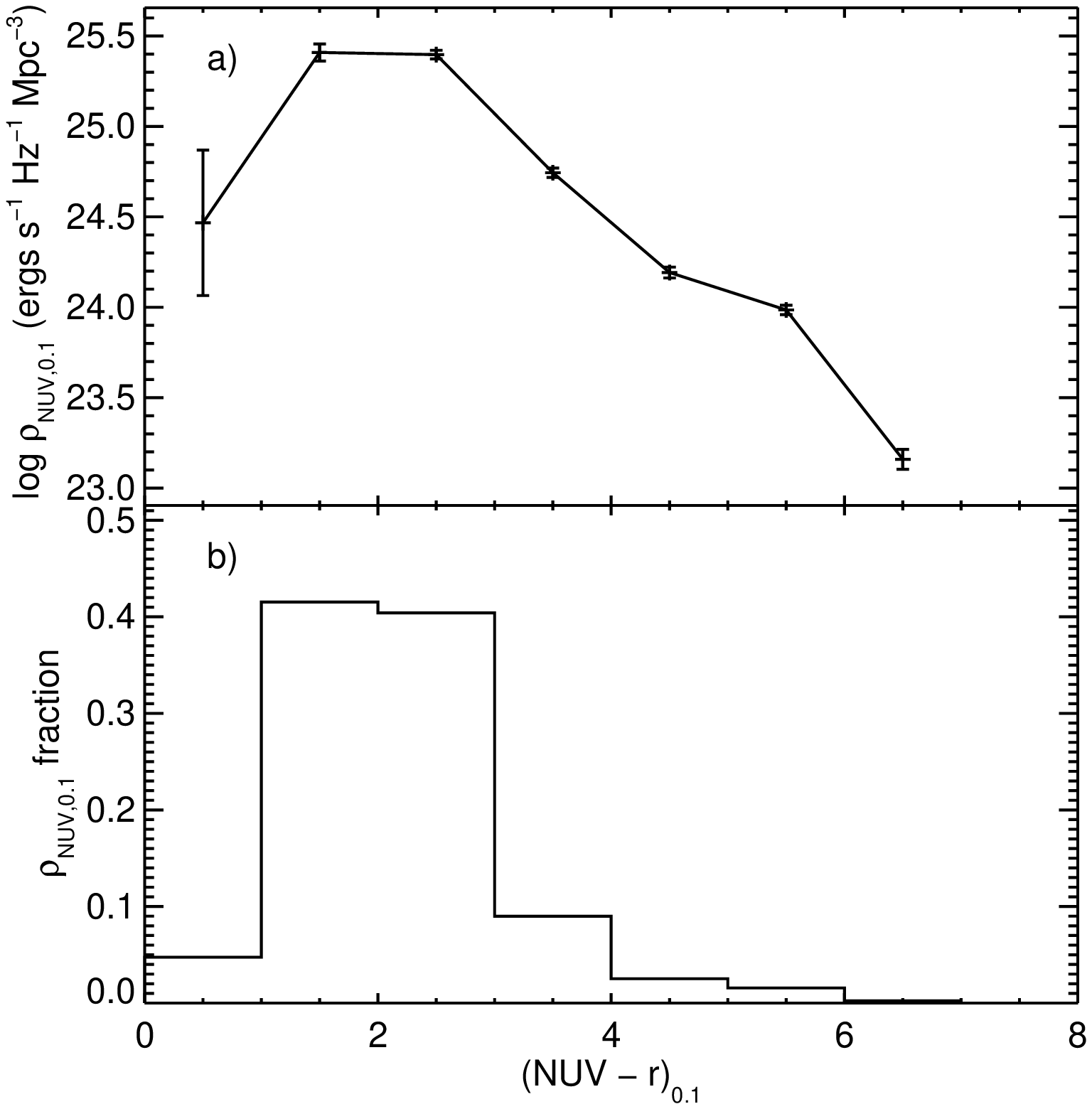}
\caption{(a) The $NUV_{0.1}$ band luminosity density $\rho_{NUV,0.1}$ as a function of $(NUV-r)_{0.1}$ color calculated by integrating the luminosity functions using the best-fit Schechter function parameters listed in Table \ref{schechter_nuv}. (b) The fraction of the total luminosity density contributed from each color bin.\label{lumden_uv}}
\end{figure}

Similar to the $r$-band, we have computed the $NUV$ luminosity density $\rho_{NUV,0.1}$ within each color bin. As before, the luminosity function for the bluest galaxies is near a slope $-2$ where its integral diverges. Therefore, for the luminosity density in this bin, we have only integrated the luminosity density down to $M_{NUV,0.1}=-12$. The value of the luminosity density as a function of color is plotted in the top panel of Figure \ref{lumden_uv} while the bottom panel shows the fraction of the total luminosity density contributed by the galaxies in each color bin. As would be expected, the $NUV$ luminosity density is dominated by the blue sequence galaxies. Specifically, $\approx80\%$ of $\rho_{NUV,0.1}$ is coming from galaxies with colors in the range $1<(NUV-r)_{0.1}<3$. The bluest galaxies, those with $0<(NUV-r)_{0.1}<1$, only contribute $\approx6\%$ to the $NUV$ luminosity density. Galaxies as blue as $(NUV-r)_{0.1}\sim0$ are difficult to produce using models with smoothly decling star formation histories. Such very blue galaxies can be reproduced by models with a star formation burst lasting $10-100$ Myr, with little dust and involving a significant fraction of the mass of the galaxy \citep{treyer98}. Clearly, such dust-free starburst galaxies are relatively rare in the local universe and do not contribute much to the total UV luminosity density. The blue sequence could in principle include galaxies undergoing large starbursts that have relatively large extinctions. However, we argue in \S3.5 below that the bulk of the galaxies in the blue sequence do not have such extreme dust attenuation.

Adding up the total luminosity density for galaxies of all colors, we obtain a value of $\log{\rho_{NUV,0.1}} = 25.791 \pm 0.029$ ergs s$^{-1}$ Hz$^{-1}$ Mpc$^{-3}$, a value consistent to within the uncertainties with the value determined by \citet{wyder05} from an UV-selected sample.

\subsection{Comparison of $(NUV-r)_{0.1}$ with $(u-r)_{0.1}$}

Although there are a number of qualitative similarities between the $(NUV-r)_{0.1}$ CMD presented here and the $(u-r)$ diagram from \citet{baldry04}, there are a few notable differences. As already shown in \S3.2, the $(NUV-r)_{0.1}$ color distributions are not well fit by a double Gaussian function, in contrast to the $(u-r)$ distributions. There is an excess of galaxies in  between the two sequences above that predicted by the double Gaussian. Moreover, the blue sequence appears to merge with the red sequence for the most luminous galaxies in the $(u-r)$ diagram while there is still a distinct blue peak visible in Figure \ref{cmd_nuv_colordist} up to $M_{r,0.1} =-22.75$. In addition, the separation between the blue and red sequences, compared to their widths is larger at all luminosities than in the $(u-r)$ distributions.

\begin{figure}
\includegraphics*[width=5in,height=5in]{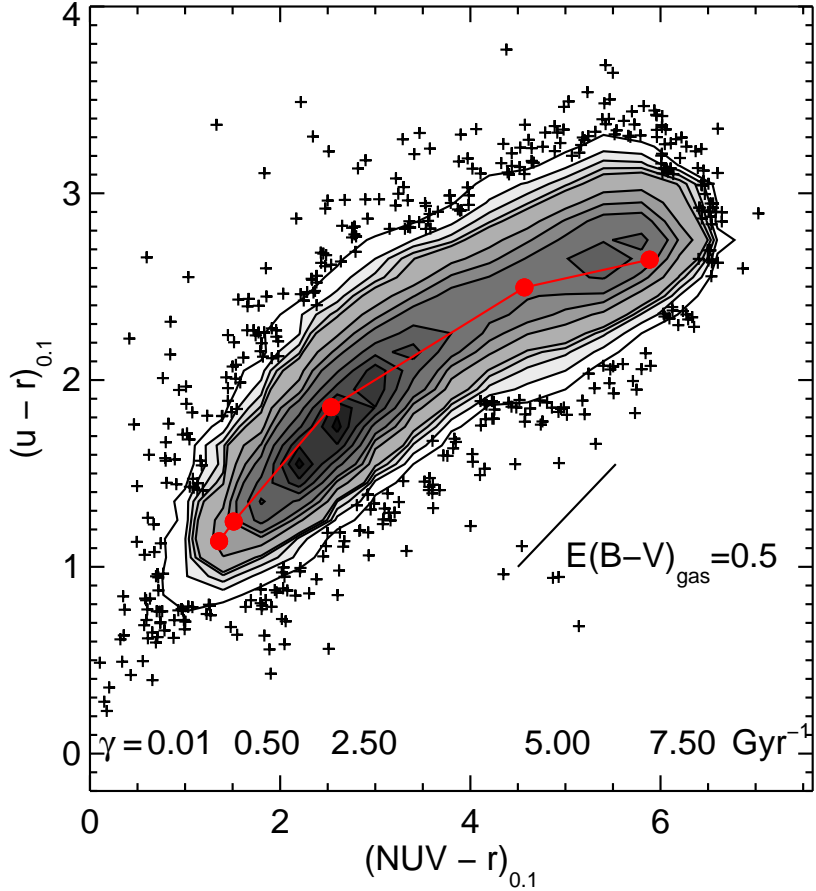}
\caption{The number of galaxies (not weighted by $1/V_{max}$) as a
function $(u-r)_{0.1}$ and $(NUV-r)_{0.1}$. These are the same
galaxies plotted in Figure \ref{cmd_nuv_points} except that galaxies
with $u$-band errors greater than 0.3 mag were excluded. The data are plotted
as contours where the density of galaxies is high and as individual points where
it is low. The solid
red circles are \citet{bruzual03} models at an age of 13 Gyr with no
dust, solar metallicity, and exponentially declining star formation
histories. The time constant in units of Gyr$^{-1}$ in the star
formation history for each model is indicated at the bottom of the
figure. The black line indicates the reddening vector in this diagram for $E(B-V)_{gas}=0.5$ and
assuming the \citet{calzetti00} attenuation law.\label{ur}}
\end{figure}

The reason for these differences is relatively easy to understand and related to the greater sensitivity of the $NUV$ band to changes in the recent star formation rate. To illustrate this point, we compare directly the $(NUV-r)_{0.1}$ and $(u-r)_{0.1}$ colors for our sample in Figure \ref{ur}. When generating this figure, we excluded galaxies with $u$-band photometric errors larger than 0.3 mag. For blue galaxies, the two colors are very well-correlated with a slope of $\Delta (u-r)_{0.1}/\Delta (NUV-r)_{0.1} \sim 0.5$. However for galaxies with colors redder than $(NUV-r)_{0.1}\approx3.5$, there is a change in slope and the $(u-r)_{0.1}$ color begins to increase less quickly with $(NUV-r)_{0.1}$ than for bluer $(u-r)_{0.1}$ colors. As a result, galaxies that are on the red sequence in the $(u-r)_{0.1}$ CMD tend to be more spread out in color in the $(NUV-r)_{0.1}$ diagram. 

This behavior is basically that predicted based upon simple galaxy models. In Figure \ref{ur}, we overplot as red circles the locations of a few \citet{bruzual03} models with an age of 13 Gyr, no dust and solar metallicity. The models are plotted for exponentially declining star formation histories with five values of the time constant $\gamma$ in the range $0.01-7.5$ Gyr$^{-1}$. These models are capable of reproducing the locus of data points in Figure \ref{ur}, and in particular the change in slope for $(NUV-r)_{0.1}>3.5$. The main resulting difference in the observed CMDs is a greater sensitivity of the UV to small changes in the recent star formation rate, especially relevant to galaxies on the red sequence or in between the two sequences. For reference, the solid black line in the figure indicates the reddening vector in this diagram corresponding to a reddening in the ionized gas of $E(B-V)_{gas}=0.5$ mag, assuming the attenuation law from \citet{calzetti00}. The reddening vector lies nearly parallel to the blue sequence, meaning that dust would tend to simply move galaxies along the blue sequence in this diagram. 

\subsection{Correcting for Dust}

One of the most important obstacles in interpreting the galaxy CMD is the fact that the UV minus optical color of a galaxy is not only affected by the galaxy's star formation history but by other physical parameters, primarily the amount of dust and the metallicity. We would like to understand how much of the variation in color with luminosity that we observe is due to which of these physical parameters. The most reliable method for determining the UV attenuation is the far-infrared (FIR) to UV flux ratio because it is almost independent of the age of the stellar population, the dust geometry, or intrinsic dust properties \citep{gordon00}. However, the vast majority of our sample have no FIR data available. Thus, we are forced to use more indirect methods. We have estimated the effects of dust on the colors of galaxies in our sample using two methods, the first using the Balmer lines, and the second using the empirical dust-SFH-color relation derived by \citet{johnson06}. 

\subsubsection{Balmer decrement}

We have used the Hydrogen Balmer line fluxes measured in the SDSS fiber spectra by \citet{tremonti04} to estimate the attenuation for our sample. The intrinsic ratio of H$\alpha$ to H$\beta$ flux is relatively independent of the physical conditions within \ion{H}{2} regions and has a value of $R_{\alpha\beta,0}=2.87$ \citep{osterbrock89}. For those galaxies with H$\alpha$ and H$\beta$ emission lines detected, we have computed a reddening in the ionized gas $E_{gas}(B-V)$ using 
\begin{equation}
E_{gas}(B-V) = \frac{2.5 \log{(R_{\alpha \beta} / R_{\alpha \beta, 0})}}{k({\rm H}\beta) - k({\rm H}\alpha)},
\end{equation}
where $k({\rm H} \beta) - k({\rm H} \alpha) = 1.163$ for the Galactic extinction law of \citet{cardelli89} with $R_V=3.1$. We further computed the attenuation in the $NUV$ and $r$ bands using the \citet{calzetti00} attentuation law, where $A_{NUV} = 3.63 E_{gas}(B-V)$ and $A_r = 1.57 E_{gas}(B-V)$. We have used the line flux uncertainties to compute a statistical uncertainty $\delta A_{NUV}$ in the resulting attenuation. For those galaxies with no H$\alpha$ or H$\beta$ lines, $A_{NUV} < 0$, or $\delta A_{NUV} < 0.5$ mag, we have assumed $E_{gas}(B-V) = 0$. For the remaining galaxies, the median $NUV$ attenuation is $A_{NUV} = 1.2$ mag with 99\% of the galaxies lying in the range $A_{NUV} = 0-3$ mag. 

There are a few caveats to bear in mind when estimating the attenuation using the Balmer lines. The relationship between the $R_{\alpha \beta}$ measured in the $3\arcsec$ diameter fiber spectra and the value for the galaxy as a whole is likely complicated. For many star-forming galaxies, the metallicity becomes lower with radius, which we would expect to lead to a decrease in the dust attenuation with radius as well. In these cases, the UV attenuation within the fiber would be higher than for the galaxy as a whole. On the other hand, some galaxies, for example with a strong bulge, have most of their star formation occurring in the outer disk of the galaxy. For these cases where there would be weak or no emission lines in the fiber spectra, the integrated UV attenuation would be underestimated. Finally, the attenuation law of \citet{calzetti00} was calibrated using observations of starburst galaxies. It has been shown that the relation ship between the UV attenuation and UV spectral slope for less active, more "normal" star-forming galaxies is smaller than predicted by the starburst results \citep{bell02, kong04, seibert05, buat05}. Whether the attenuation determined from the Balmer decrement and the \citet{calzetti00} law lead to similar overestimates of the attenuation remains unclear. 

\begin{figure}
\includegraphics*[width=5in,height=5in]{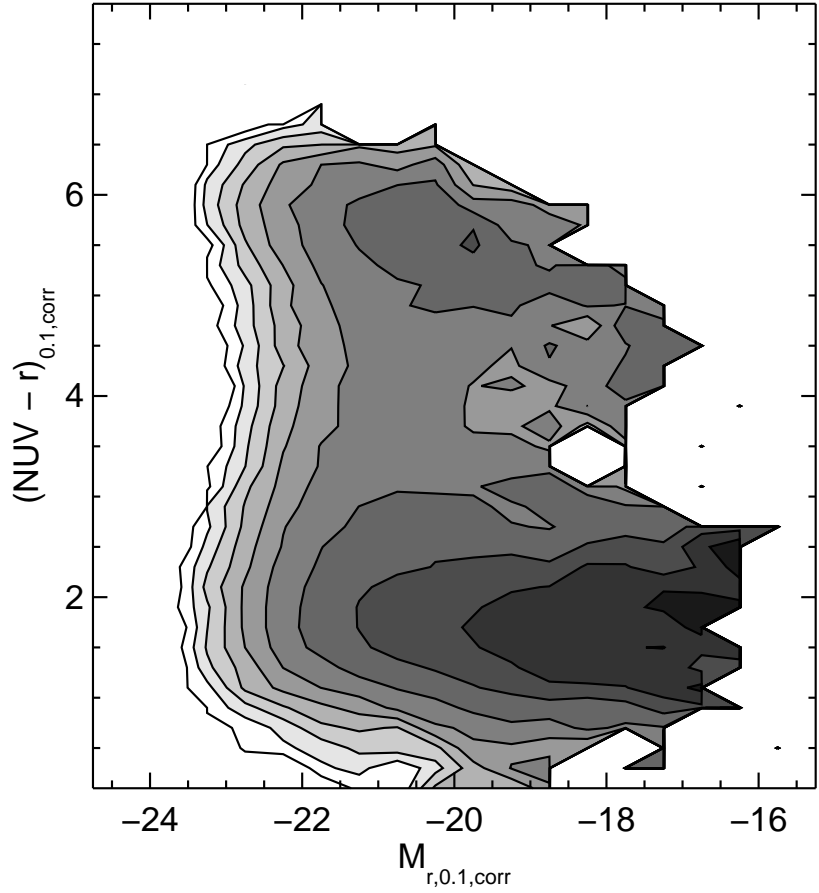}
\caption{The number density of galaxies as a function of dust-corrected color $(NUV-r)_{0.1,corr}$ and absolute magnitude $M_{r,0.1}$. The $NUV$ and $r$ band attenuation has been calculated from the Balmer decrement and assuming the extinction law of \citet{calzetti94} as described in \S3.5.1. The contours are spaced logarithmically from $10^{-5.5}$ to $10^{-2.3}$ Mpc$^{-3}$ mag$^{-2}$. \label{cmd_nuv_extcorrbalmer_vmax}}
\end{figure}

The dust-corrected CMD is plotted in Figure \ref{cmd_nuv_extcorrbalmer_vmax}. The color of the faint end of the blue sequence changes little since the reddening for these galaxies is small. The reddening increases with luminosity such that the overall trend of color with luminosity disappears while leaving the dispersion of the blue sequence relatively unchanged. The dust-corrected color of the peak of the blue sequence has a value of $(NUV-r)_{0.1} \approx 1.7$.

\subsubsection{The dust-SFH-color relation}

By combining a spectroscopic measure of the star formation history relatively insensitive to dust, such as the $D_n(4000)$ index \citep[e.g.,][]{kauffmann03a}, with UV, optical, and FIR fluxes, it is possible to separate the effects of dust and star formation history on the colors of galaxies. \citet{johnson06} have used a sub-sample of SDSS galaxies with both UV measurements from {\it GALEX} and FIR fluxes from {\it Spitzer} to determine the UV attenuation, as measured by the FIR to UV ratio, as a function of $D_n(4000)$ and color. We have utilized the fits given by \citet{johnson06} to estimate the UV attenuation for our entire sample. Specifically, we calculated the $FUV$ attenuation $A_{FUV}$ as a function of $D_n(4000)$ and $(NUV-r)_{0.1}$ color from the fits in Table 1 of \citet{johnson06} as
\begin{equation}
A_{FUV} = 1.27 - 1.56\{D_n(4000)-1.25\} + 1.35\{(NUV-r)_{0.1}-2\} - 1.24\{D_n(4000)-1.25\} \{(NUV-r)_{0.1}-2)\}.
\label{extinction_eqn}
\end{equation}
We calculated the $NUV$ and $r$ attenuation using equation (\ref{extinction_eqn}) and assuming $A_{NUV} = 0.81 A_{FUV}$ and $A_r = 0.35 A_{FUV}$, derived from the \citet{calzetti00} attenuation curve.

\begin{figure}
\includegraphics*[width=5in,height=5in]{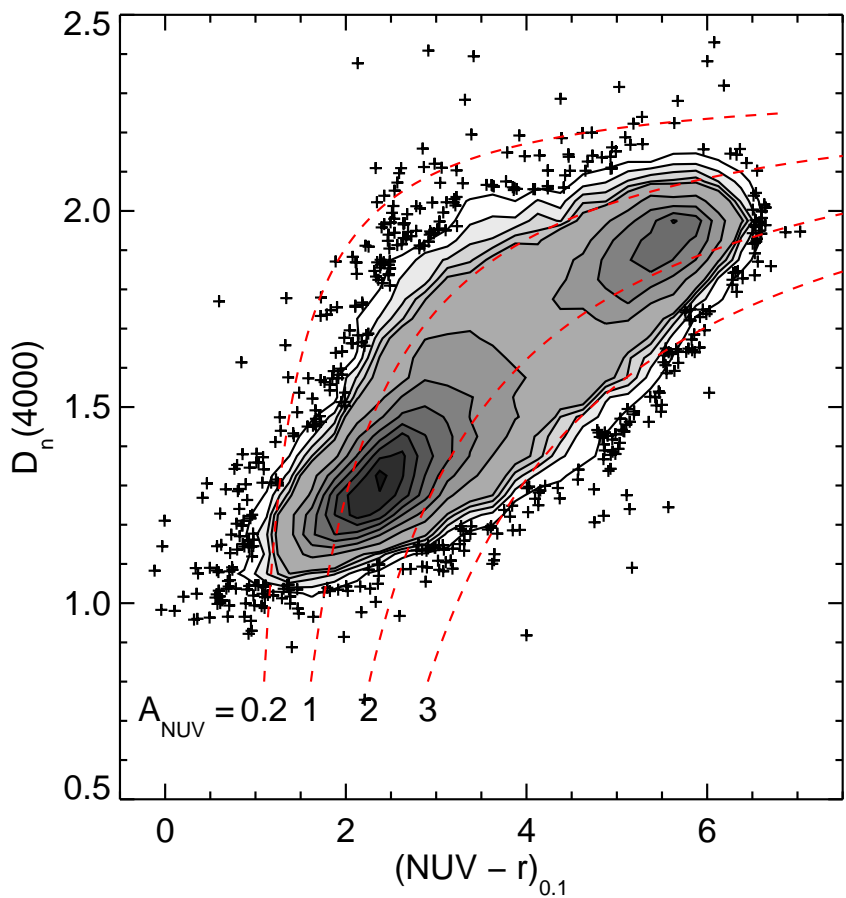}
\caption{The number of galaxies (not weighted by $1/V_{max}$) as a
function of $(NUV-r)_{0.1}$ and $D_n(4000)$ from the SDSS spectra. 
The data are plotted as contours where the density of galaxies is high 
and as individual points where it is low. The
dashed curves indicate lines of constant $NUV$ extinction ranging from
0.2 to 3 magnitudes as derived from the fits in \citet{johnson06}. \label{d4000}}
\end{figure}

In order to illustrate the range of $NUV$ attenuations in our sample, in Figure \ref{d4000} we plot the  $(NUV-r)_{0.1}$ color as a function of $D_n(4000)$ from the SDSS spectra. The contours in the figure reflect the number of galaxies in the sample and have not been weighted by $1/V_{max}$. Overall, the $(NUV-r)_{0.1}$ color and the $D_n(4000)$ index are correlated with redder galaxies having on average larger $D_n(4000)$ values, indicating older stellar populations, although with a relatively large spread. The dashed red lines plotted over the data in Figure \ref{d4000} are lines of constant $A_{NUV}$ using equation (\ref{extinction_eqn}) . The bulk of the blue sequence galaxies in our sample, with absolute magnitudes near $M^*$, have $A_{NUV}=1-2$ mag, with the mode of the distribution at $A_{NUV}\approx1.3$ mag. The attenuation decreases on average with decreasing color and reaches values near zero for the bluest galaxies. Since the distribution of galaxy colors along the blue sequence do not lie perpendicular to lines of constant attenuation in Figure \ref{d4000}, the variation in color along the blue sequence is not entirely due to dust alone $-$ the star formation history and metallicity also play a role.

While the $(NUV-r)_{0.1}$ color is relatively tightly correlated for blue and red sequence galaxies, sources with intermediate colors tend to exhbibit a larger spread in $D_n(4000)$. Indeed, galaxies with these intermediate colors tend to have a large range of attenuation ranging from one to three magnitudes. This would indicate that some galaxies are located in between the two sequence simply because they are star-forming galaxies with a lot of dust reddening  while other galaxies are located there due to their older average stellar populations. 

In addition to dust and star formation history, it is important to keep in mind that aperture effects can play an important role, particularly for galaxies in between the two sequences because the $D_n(4000)$ measurements only sample the inner $3\arcsec$ of each galaxy. For galaxies with strong bulges, for example, the SDSS spectra sample mostly the bulge resulting in an old value for $D_n(4000)$ whereas the color would tend to indicate a somewhat younger age since it measures the flux from both the bulge and any recent star formation in the disk. This may explain the population of galaxies in Figure \ref{d4000} that have values of $D_n(4000)$ that would place them on the red sequence but with colors that place them in between the red and blue sequences.  However, as the sample used by \citet{johnson06} is drawn from the SDSS, their fits should inherently account for this aperture correction on average.

The relation in equation (\ref{extinction_eqn}) is an empirical description of the correlations observed in the joint SDSS-{\it GALEX}-{\it Spitzer} sample analyzed by \citet{johnson06}.  As such, it is difficult to interpret directly in terms of physical variables. Presumably, it is the galaxies' star formation history which varies along each line of constant UV attenuation shown in Figure \ref{d4000} although the lines also implicitly account for the fact that the $D_n(4000$) index is measured only within the central regions of each galaxy. In addition, the analysis of \citet{johnson06} assumes a very simple prescription for converting the FIR/UV ratio to an UV attenuation and does not take into account heating due to an older stellar population. In such cases, particularly relevant for red sequence galaxies with little or no recent star formation, the UV and FIR emission may become decoupled. Indeed, the residuals from the fit in equation (\ref{extinction_eqn})  are larger for red galaxies in the \citet{johnson06} sample.

Another important caveat to bear in mind when interpreting the lines of constant $A_{NUV}$ in Figure \ref{d4000} is the cuts imposed to define the sample used in determining the fit in equation (\ref{extinction_eqn}). The sample used by \citet{johnson06} was selected by requiring galaxies to be detected both in the UV by {\it GALEX} and at $24\micron$ by {\it Spitzer}. Thus, their sample will not include quiescent early type galaxies with no recent star formation. As can be seen in Figure \ref{d4000}, the typical attenuation predicted for red sequence galaxies is $A_{NUV} \sim 1.5$ mag. This is most likely  an overestimate in many cases since many of these red sequence galaxies likely do not contain any residual star formation and are simply red due to their older stellar populations. Therefore, care must be taken in interpreting the UV attenuation for red sequence galaxies in our sample. 

\begin{figure}
\includegraphics*[width=5in,height=5in]{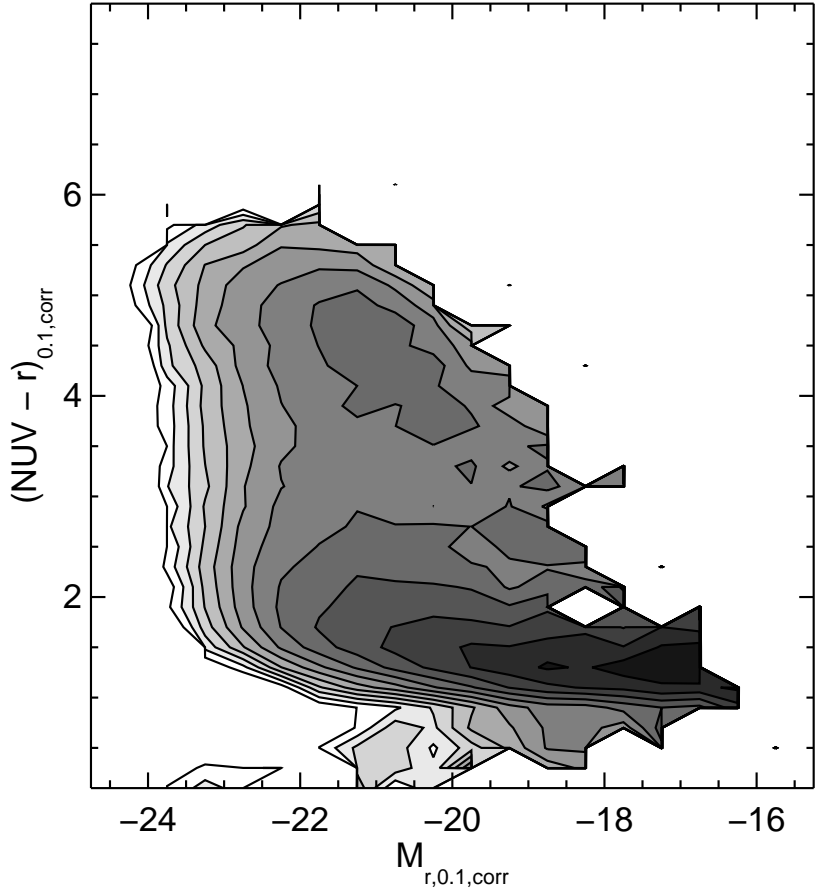}
\caption{The volume density of galaxies as a function of dust-corrrected $(NUV-r)_{0.1,corr}$ and $M_{r,0.1,corr}$. The $FUV$ attenuation was derived using equation (\ref{extinction_eqn}) in the text based upon the analysis of \citet{johnson06}. The corresponding values of $A_{NUV}$ and $A_r$ were calculated using $A_{NUV}=0.81A_{FUV}$ and $A_r=0.35A_{FUV}$ from the \citet{calzetti00} attenuation law. The contours are spaced logarithmically from $10^{-5.5}$ to $10^{-2.0}$ Mpc$^{-3}$ mag$^{-2}$. \label{cmd_nuv_extcorr_vmax}}
\end{figure}

The $(NUV-r)_{0.1}$ vs. $M_{r,0.1}$ diagram corrected for dust using the dust-SFH-color relation is shown in Figure \ref{cmd_nuv_extcorr_vmax}. In this corrected diagram the ridge line of the blue sequence is shifted towards the blue compared to the uncorrected diagram while the width has become significantly narrower, particularly at the faint end. In contrast to the Balmer dust corrected CMD, a trend of color with absolute magnitude remains. Whereas the ridge line of the blue sequence in the uncorrected CMD increases from about $(NUV-r)_{0.1}\approx 1.8$ to 3 with increasing luminosity, the blue sequence in the dust corrected diagram increases from $(NUV-r)_{0.1}\approx 1.3$ to $2.2$. Thus, roughly a quarter of the change in color of the blue sequence with luminosity is due to dust. The red sequence in the corrected diagram also shifts to the blue but has a wider spread in color compared to the uncorrected CMD. As we have already argued, it is likely that the attenuation estimated for part of the red sequence using equation (\ref{extinction_eqn}) has been overestimated. In addition, a significant number of galaxies with intermediate colors remains.

Despite this caveat for non-star-forming galaxies, we prefer the relations computed by \citet{johnson06} as the best measure of the UV attenuation for our sample because this method is based upon the total FIR/UV ratio, which is the most direct measure of the UV light absorbed \citep[e.g.][]{gordon00}. Deriving an UV attenuation from the Balmer lines requires assuming an attenuation law to convert from the extinction in the Balmer lines to that in the UV in addition to the uncertainty associated with only having measurements in the central $3 \arcsec$ of each galaxy. Although \citet{calzetti94} showed that $E_{stars}(B-V) = 0.44 E_{gas}(B-V)$ for starburst galaxies, it remains unclear whether this relation pertains to more "normal" galaxies. 

Besides attenuation due to dust and the star formation history, the colors of galaxies can also be affected by the metallicity. We have estimated the effect of metallicity on the colors of star-forming galaxies using the models of \citet{bruzual03}. For galaxies at an age of 12.6 Gyr, no extinction, and exponentially declining star formation histories with time constant $\gamma < 4$ Gyr$^{-1}$, the $(NUV-r)$ color varies by 0.2 mag for metallicities between 0.4 and 2.5 times the solar abundance. For such currently star-forming galaxies, the $(NUV-r)$ color becomes bluer with increasing metallicity. This some what counterintuitive result is due to the variation in the ratio of blue to red supergiant stars as a function of metallicity \citep{bruzual03}. Since metallicity tends to increase with galaxy luminosity \citep[e.g.,][]{tremonti04}, the increase in color along the blue sequence would be expected to be decreased somewhat due to this effect. Thus, metallicity variation does not account for the remaining change in color along the blue sequence after correcting for dust attenuation.

\subsection{Specific Star Formation Rates as a Function of Stellar Mass}

Although we do not have access to the detailed star formation history of each of the galaxies in our sample, we can try to place some basic constraints on their evolution. One of the most basic parameters that we may hope to use to constrain the star formation history is the specific star formation rate, or the star formation rate per stellar mass $SFR/M^*$. We have made estimates of specific star formation rates for our sample as follows. For the stellar masses, we have made use of the estimates already included as a part of the MPA/JHU values-added catalogs. As described in detail by \citet{kauffmann03a}, the stellar masses were determined from the $z$-band luminosities with the mass to light ratio constrained by their model fits to the $D_n(4000)$ and $H\delta_A$ spectroscopic indices. The masses assume a \citet{kroupa01} stellar initial mass function (IMF). We have estimated a star formation rate SFR in M$_{\sun}$ yr$^{-1}$ for each galaxy using its dust corrected $NUV$ luminosity. We have relied upon the relation from \citet{kennicutt98} converted to the \citet{kroupa01} IMF: $SFR = 1.0\times10^{-28} L_{\nu,NUV}$,  where $L_{\nu,NUV}$ is the $NUV$ luminosity in units of ergs s$^{-1}$ Hz$^{-1}$ and SFR is the star formation rate in $M_{\odot}$ yr$^{-1}$.

An alternative method for calculating stellar masses was presented by \citet{bell03}. Based upon fits of models to galaxies with optical and near-IR photometry, \citet{bell03} determined relations between the stellar mass-to-light ratio $(M_{*}/L)$ and color for various bands. Using their relation between $(M_{*}/L)_r$ and $(g-r)$, we have compared the resulting stellar masses with those from \citet{kauffmann03a}. The stellar masses agree on average for the most massive galaxies while those for lower mass galaxies are larger using the \citet{bell03} relations. The difference in mass increases steadily below $10^{10}$ M$_{\sun}$, reaching a difference of $\approx 0.4$ dex on average at $10^8$ M$_{\sun}$. Throughout the remainder of this discussion, we assume the stellar masses from \citet{kauffmann03a} but note that the specific SFRs for the lowest mass galaxies are likely more uncertain since the origin of this difference between the two mass calibrations is not clear.

\begin{figure}
\includegraphics*[width=5in,height=5in]{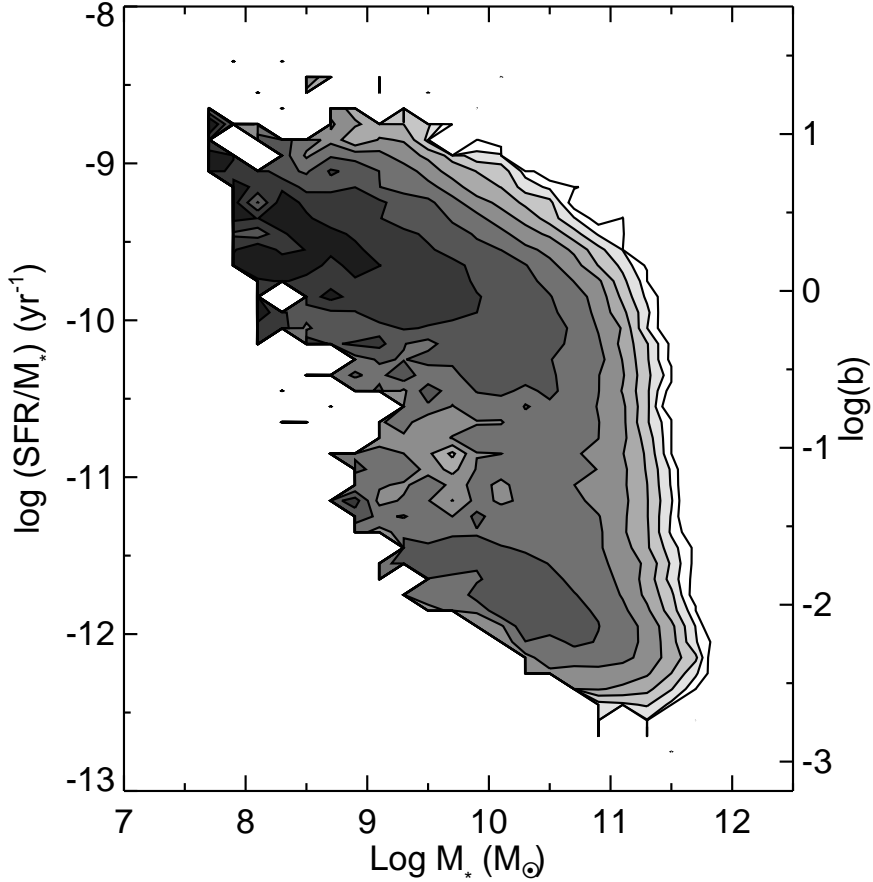}
\caption{The volume density of galaxies as a function of specific SFR and stellar mass $M_*$. The SFR has been calculated from the $NUV$ luminosity, corrected for dust using the Balmer lines measured from the SDSS fiber spectra as described in the text. The density was calculated in bins 0.2 dex wide in mass and 0.1 dex wide in specific star formation rate. The contours are spaced logarithmically from $10^{-5}$ to $10^{-1.9}$ Mpc$^{-3}$ dex$^{-2}$. We converted the $NUV$ luminosities
to SFRs using the conversion factor given in \citet{kennicutt98} after
applying a correction factor to convert to the \citet{kroupa01} IMF
used to calulcute the stellar masses. The axis on the right hand side gives the value of
the logarithm of the ratio of current to past averaged SFR, $\log{(b)}$, calculated as described in the text.
\label{ssfr_balmer}}
\end{figure}

\begin{figure}
\includegraphics*[width=5in,height=5in]{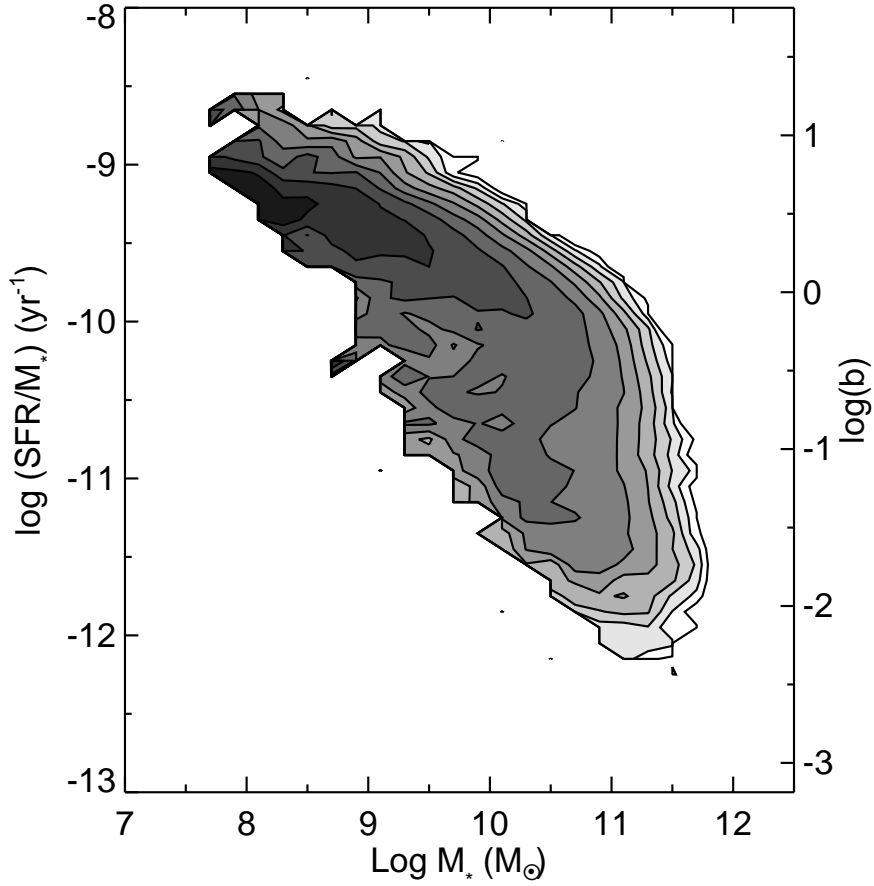}
\caption{Similar to Figure \ref{ssfr_balmer}, except that the SFR has been calculated from the $NUV$ luminosity using the dust-SFH-color relation from \citet{johnson06}. The contours are spaced logarithmically from $10^{-5}$ to $10^{-1.5}$ Mpc$^{-3}$ dex$^{-2}$.
\label{ssfr_johnson}}
\end{figure}

The number density of galaxies determined using the $V_{max}$ weighting is shown in Figures \ref{ssfr_balmer} and \ref{ssfr_johnson} for the two different dust corrections discussed in \S3.5. In Figure \ref{ssfr_balmer}, the SFR was calculated from the $NUV$ luminosity corrected for dust using the Balmer decrement, whereas the SFR used in Figure \ref{ssfr_balmer} was determined using the dust-SFR-color relation of \citet{johnson06}. The overall trends are similar in both diagrams. The specific SFR is correlated strongly with the stellar mass, as previous studies have shown \citep[e.g.][]{brinchmann04}. The specific SFR decreases gradually with increasing stellar mass. Above a stellar mass of $\sim10^{10.5}$ M$_{\sun}$, galaxies with very low specific SFRs begin to dominate, although they are present in smaller numbers well below this mass. This is partially a selection effect since our sample would be biased against low mass galaxies with low specific SFRs. Indeed, the low mass limit in Figures \ref{ssfr_balmer} and \ref{ssfr_johnson} increases as a function of mass due to the UV selection limit of our sample. 

There are some notable differences between the specific SFRs in Figures \ref{ssfr_balmer} and \ref{ssfr_johnson}. The width of the blue sequence is substantially broader when using the Balmer  decrement dust correction. In both diagrams the peak specific SFR in both diagrams is $\approx 10^{-10.3}$ yr$^{-1}$ at the high mass end of the blue sequence at $10^{11}$ M$_{\sun}$. The specific SFR increases somewhat less with decreasing mass for the Balmer decrement dust correction than for the \citet{johnson06} dust correction. At a stellar mass of $10^{8.5}$ M$_{\sun}$, the peak specific SFR is $\approx 10^{-9.6}$ yr$^{-1}$ for the Balmer decrement dust correction and $\approx 10^{-9.3}$ with the \citet{johnson06} dust correction. 

There are larger differences in the specific SFRs derived from the two dust corrections for red sequence galaxies. The specific SFR is a factor of $\sim 10$ lower for the Balmer decrement dust correction. Determining specific SFRs for red sequence galaxies is problematic in both methods. The Balmer line dust correction underestimates the amount of dust absorption in cases where most of the star formation is taking place in the outer parts of a galaxy that are not sampled by the SDSS fiber spectra. On the other hand, the \citet{johnson06} method overestimates the dust absorption in some red sequence galaxies. In particular, dust heating by older stellar populations or AGN can contribute to the FIR luminosity to a higher degree in red sequence galaxies. In addition, the galaxies used to derive the correlations in \citet{johnson06} were all detected at $24\micron$, which would bias the sample against truly quiescent galaxies with very little dust. Regardless of  the dust correction applied, SFRs derived from the $NUV$ luminosity for red sequence galaxies are problematic because the $NUV$ band can include light from older stellar populations \citep{yi05, rich05}. In those cases, the $NUV$ luminosity will overestimate the recent SFR. For these reasons, caution must be used when interpreting the specific SFRs for red sequence galaxies. Figures \ref{ssfr_balmer} and \ref{ssfr_johnson} are most useful for investigating the star formation histories of blue sequence galaxies. As we have argued in the previous section, we prefer the \citet{johnson06} dust correction and thus restrict our discussion to Figure \ref{ssfr_johnson} through out the remainder of the paper.

With a few additional assumptions, the specific SFR can be related to the ratio of the present to past average SFR as $b = SFR/<SFR> = (SFR/M^*) T R$, where $T$ is the age of the galaxy and $R$ is the fraction of the mass formed over the galaxy's lifetime that does not eventually get returned to the ISM or IGM \citep{brinchmann04}. A typical value of $R$ is $\approx 0.5$ \citep{brinchmann04}. Assuming that all galaxies started forming stars shortly after the Big Bang, we set $T=13$ Gyr. For these choices of $T$ and $R$, a galaxy with $b=1$, i.e. a constant SFR, would have a specific SFR of $10^{-9.8}$ yr$^{-1}$. Thus galaxies with specific SFRs larger than this value would have current SFRs above their average while those with specific SFRs $<10^{-9.8}$ yr$^{-1}$ would have current SFRs  below their lifetime average. The right hand y-axes of Figures \ref{ssfr_balmer} and \ref{ssfr_johnson} gives the corresponding values of $\log{(b)}$. Under these assumptions, blue sequence galaxies with $M_* < 10^{10}$ M$_{\sun}$ would have star formation histories increasing moderately with time, with more massive blue sequence galaxies having formed their stars at somewhat declining rate with time. As noted earlier when discussing the CMD itself, galaxies undergoing extreme starbursts, i.e. objects with $\log{(b)} > 1$, are rare in the local universe and do not contribute significantly to the SFR density.

Quantitatively, the specific star formation rates estimated here are somewhat larger than those determined by \citet{brinchmann04} from their analysis of the SDSS emission lines, especially for lower mass galaxies. The origin of this difference is not clear. The SFRs estimated by \citet{brinchmann04} were derived largely from the emission lines coming from \ion{H}{2} regions within the SDSS fiber. Since stars capable of producing an \ion{H}{2} region are very massive stars with lifetimes up to $\sim 10^7$ yr, compared to stars with lifetimes up to $\sim 10^8$ yr that dominate the UV emission, differences between UV and optical emission line based SFRs could be due to short time scale variations in the SFRs of galaxies. This could in principle be more important for lower mass galaxies where stochastic fluctuations in the SFR would be expected as a fraction of the total mass. In addition to this fundamental difference, part of the difference could be due to uncertainties in our method for correcting for dust attenuation or to uncertainties in the aperture corrections applied to the SDSS measurements. A more detailed comparison of the SFRs from \citet{brinchmann04} and {\it GALEX} UV determinations is presented elsewhere \citep{treyer07}.

\section{Discussion}

Due to the greater sensitivity of the UV minus optical colors to very low specific star formation rates as compared to optical colors, our galaxy color-magnitude diagrams have revealed a few new features. Similar to optical results, our CMD exhibits prominent blue and red sequences. However, whereas in the $(u-r)$ CMD \citep[e.g.,][]{baldry04}, the blue sequence appears to merge with the red sequence at the luminous end, in the $(NUV-r)_{0.1}$ CMD, the blue sequence remains as a separate peak in the color distribution up to $M_{r,0.1} \simeq -23$. Also in the optical CMDs \citep{baldry04, balogh04}, the color distribution in each absolute magnitude bin is well fit by the sum of two Gaussians. This color bimodality naturally leads to an interpretation in which there are simply two populations of galaxies with nothing in between. As we have shown in Figure \ref{cmd_nuv_colordist}, the color distributions as a function of $M_{r,0.1}$ are not well fit by a double Gaussian function due to an excess of galaxies in between the red and blue peaks. These galaxies in the "green valley" between the red and blue peaks would suggest that the properties of galaxies are not strictly bimodal and that there really exists more of a continuum of properties between star-forming and quiescent galaxies. 

Besides the colors, we have also shown that the $r_{0.1}$-band luminosity function shape varies systematically with color. While there is a steady increase in the faint end power law exponent $\alpha$ with color across the blue peak, the slope levels off in the "green valley." In addition, the decrease in the value of the characteristic magnitude $M^{*}_{r,0.1}$ appears to level off at intermediate colors as well. This change in the luminosity function shape at intermediate colors would also suggest that the "green valley" galaxies coincide with a physical difference in the galaxy populations.

Interpreting the change in the peak color of each sequence with luminosity is complicated by the fact that a galaxy's color is affected not only by its star formation history, but also its metallicity and dust content. For morphologically selected early type galaxies, there is little dust and the change in color is due to some combination of metallicity and age of the stellar populations. While the red sequence, as defined by the CMD, contains galaxies with a range of morphologies, the dominant effects are still likely the age and metallicity, although a significant fraction of early-type galaxies appear to have some residual star formation \citep{yi05}. 

On the other hand, for blue sequence galaxies, we know that their colors are affected both by dust attenuation and their SFH, with metallicity playing a minor role. We have attempted to determine the effects of dust in two ways. Using the Balmer decrement as observed within the SDSS fiber spectra in conjunction with the \citet{calzetti00} attenuation law, we found that the trend of the peak color with luminosity disappears, which would suggest that the average color along the blue sequence is due mostly to dust with the SFH primarily responsible for the dispersion in the color at fixed luminosity. 

Our second method to correct for dust reddening and attenuation relies upon the empirical correlations derived by \citet{johnson06}. Applying these fits to our entire sample, we found that only about one quarter of the change in color along the blue sequence is due to dust. In this case the dispersion of the blue sequence is small and there is a significant variation in the average color with luminosity, indicating a decrease in the average galaxy age with decreasing luminosity. Due to the requirement that the galaxies used to calibrate these relations must be detected in both the UV and at $24 \micron$, there is an inherent bias against truly quiescent galaxies with little star formation and dust. Thus, the attenuation for the red galaxies is likely overestimated. Nevertheless, we argued in \S3.5 why we prefer this dust correction.

While the detailed morphology of the CMD depends upon which dust correction is assumed, there remain galaxies in the "green valley" between the two sequences in both versions of the dust-corrected CMD. While there are indeed some galaxies with intermediate colors that are simply very reddened versions of blue sequence galaxies, many do have SFHs weighted to older ages. We would expect that as time progresses, the red sequence is building up as galaxies exhaust their gas and cease forming stars. With the data presented here we may be beginning to separate out the population undergoing this transition. Of course, it is important to note that just based upon the integrated measurements here it is not possible to infer the evolution of different structures within each galaxy. In fact, galaxies with bulges would be expected to lie at intermediate colors, where the SFH of the bulge may be quite different than that of the disk. In these cases, the integrated color would reflect the relative strengths of the bulge and disk. In addition, it is not possible just based upon the simple diagnostics employed in this paper to infer whether galaxies in the green valley are star-forming galaxies transitioning to the red sequence for the first time or whether they were already on the red sequence and underwent a burst of star formation after merging with a gas rich galaxy. The rate at which galaxies are moving from the blue to the red sequence is explored in more detail in \citet{martin07}. 

A remarkable feature of the specific SFR as a function of stellar mass in Figure \ref{ssfr_johnson} is the relatively small spread of a factor of $2-3$ in $SFR/M_*$ at a given stellar mass. It is difficult for galaxies to have a large $SFR/M_*$ ratio for an extended period of time due to exhausting the available gas supply as well as feedback from supernovae which tends to heat the gas and prevent it from forming stars. On the other hand, the energy input from supernovae can help maintain lower levels of star formation over an extended period of time as long as the halo is massive enough to retain the heated gas. 

The behavior seen in Figure \ref{ssfr_johnson} is broadly consistent with the theoretical models of \citet{cattaneo06} which succeeded in reproducing the distribution of galaxies in the optical $(u-r)$ CMD. In their models, galaxies below a critical halo mass of $10^{12}$ M$_{\sun}$ accrete gas from cold streams. This gas fuels star formation that is self-regulated due to feedback from supernovae and stellar winds. These lower mass galaxies gradually gain in mass, move up in luminosity along the blue sequence, and become redder. The gradual decrease in the specific star formation rate with stellar mass seen in Figure \ref{ssfr_johnson} would be consistent with the predictions of this model. The gradual transition to the dominance of red galaxies above a stellar mass of $\sim 10^{10.5}$ M$_{\sun}$ would correspond in this picture to the critical halo mass of $10^{12}$ M$_{\sun}$ above which star formation is quenched due to shock heating of the gas and feedback from AGN. In their model lower mass galaxies can also have their star formation quenched if they happen to become satellites within a halo above the critical mass. This tends to populate galaxies along the fainter end of the red sequence and serves to broaden somewhat the mass range over which the transition from star-forming to quiescent galaxies occurs. The most luminous blue galaxies help constrain the value of the critical halo mass $M_{shock}$ above which gas is not able to cool and form stars. While \citet{cattaneo06} have used the $(u-r)$ CMD to constrain the value of $M_{shock}$, our observations of blue sequence galaxies at slightly larger luminosities may correspond to somewhat larger values for $M_{shock}$. 

In their analysis of the evolution of galaxies in the $(U-B)$ CMD out to $z=1$, \citet{faber06} discussed the origin of red sequence galaxies and how blue sequence galaxies transition from blue to red. In their favored scenario, galaxies first grow moderately in mass along the blue sequence, then undergo a merger which halts further star formation, and finally undergo a series of gas-free mergers that move the galaxies up the red sequence. Such a scenario is broadly consistent with the results presented here. The excess of galaxies in between the two sequences would in this case be consistent with galaxies in the process of turning off their star formation and making the transition to the red sequence. As can be seen in Figures \ref{cmd_nuv_lfs} and \ref{cmd_nuv_lfs_param}, the $M_{r,0.1}$ luminosity functions for the bluest galaxies have a steeper slope and fainter $M^*$ than for galaxies in between the two sequences. If large numbers of lower mass galaxies were stopping star formation and moving towards the red sequence, we would expect to see a steeper faint end slope to the luminosity function for galaxies with intermediate and red colors. Thus, the luminosity functions are more consistent with a scenario in which the ancestors of galaxies on the red sequence are weighted towards the luminous end of the blue sequence. The bright end cut-off of the $r$-band luminosity functions is remarkably similar for galaxies with $2 < (NUV-r)_{0.1} < 5$ while there appears to be a relatively sharp jump in the bright end at $(NUV - r)_{0.1} \sim 5$. Thus, the most massive galaxies are unlikely to be the descendants of the merging of galaxies at the luminous end of the current blue sequence. Either these galaxies are the descendants of much more massive blue sequence galaxies that are not present in the nearby universe, or they are the result of "dry" mergers of smaller mass red sequence galaxies, as \citet{faber06} argued. 

\section{Summary}

We have determined the volume density of galaxies in the local universe as a function of absolute magnitude $M_{r,0.1}$ and $(NUV-r)_{0.1}$ and $(FUV-r)_{0.1}$ colors based upon a sample of galaxies observed in the UV by {\it GALEX} and with optical data from the SDSS. The galaxies in these CMDs separate into well-defined blue and red sequences that become redder with increasing luminosity. While the most luminous galaxies are on the red sequence, a separate blue peak is detectable as bright as $M_{r,0.1} \approx -23$. In contrast to CMDs relying solely on an optical color such as $(u-r)$ \citep{baldry04}, the color distribution at fixed absolute magnitude is not well fit by the sum of two Gaussians due to an excess of objects at intermediate colors between the blue and red peaks. The greater separation between the blue and red sequences is a consequence of the greater sensitivity of the UV bands to very low levels of recent star formation. The $r_{0.1}$-band luminosity function shape varies systematically with color with the faint end slope $\alpha$ gradually increasing across the blue sequence, reaching a value of $\alpha \sim -0.6$ at intermediate colors before increasing even more for the reddest galaxies. We have used these fits to the luminosity functions to derive the fraction of the luminosity density in the local universe as a function of color. Dust-free starburst galaxies with colors $(NUV-r)_{0.1}<1$ are rare in the local universe and account for only about 5\% of the $NUV_{0.1}$ luminosity density. About 80\% of the $NUV_{0.1}$ luminosity denisty is emitted by blue sequence galaxies with colors $1 < (NUV-r)_{0.1} <  3$. 

We have used both the Balmer decrement and the dust-SFH-color relation of \citet{johnson06} to estimate the effect of dust on the galaxy colors and absolute magnitudes. For the Balmer decrement method, the increase in color with luminosity along the blue sequence is due entirely to dust with the dispersion at fixed absolute magnitude relatively unchanged. On the other hand, the blue sequence color in the CMD corrected for dust using the \citet{johnson06} method does still increase with luminosity, indicating that part of this change in color is due to the star formation history and not to dust alone. We argue that we prefer the \citet{johnson06} method as it is ultimately based upon an attenuation derived from the FIR/UV ratio. Regardless of which dust correction we employ, however, a significant number of galaxies remain at colors in between the two sequences, indicating that not all of the galaxies there are simply dusty versions of blue sequence star-forming galaxies. 

We have used the $NUV_{0.1}$ luminosities corrected for dust using the \citet{johnson06} method in conjunction with the stellar masses determined by \citet{kauffmann03a} to plot the density of galaxies as a function of specific star formation rate $SFR/M_*$ and stellar mass $M_*$. The dispersion in $SFR/M_*$ is only a factor of $2-3$ at a fixed stellar mass along the blue sequence. The value of $SFR/M_*$ decreases from values of $\approx10^{-9.3}$ yr$^{-1}$ at $M_*=10^{8.5}$ M$_{\sun}$ to $\approx 10^{-10.3}$ near the tip of the blue sequence at $M_* = M^{11}$ M$_{\sun}$. Similar to previous optical results \citep{kauffmann03a, kauffmann03b}, galaxies with low specific star formation rates begin to dominate above a stellar mass of about $10^{10.5}$ M$_{\sun}$. 

In addition to the small number of galaxy properties explored here, many galaxies in our sample contain many other measurements mainly from the SDSS. In a companion paper in this volume, \citet{martin07} have estimated the mass flux of galaxies from the blue to the red sequence and have discussed some of the other properties of the galaxies in between the red and blue sequences. In another paper, \citet{schiminovich07} have investigated the correlation of morphology and other characteristics with position in the CMD. While detecting red sequence galaxies out to significant distances in the rest-frame UV is very difficult, it should be possible to use data from {\it GALEX} deep exposures in conjunction with ground-based photometry and spectra to investigate the evolution of the blue sequence with redshift. In addition, the variation of the CMD with local galaxy density should provide interesting contraints on the nature of the galaxies in between the blue and red peaks as well as models of the physical processes affecting the evolution of galaxies in the CMD.

\acknowledgments

{\it GALEX} (Galaxy Evolution Explorer) is a NASA Small Explorer, launched in April 2003.
We gratefully acknowledge NASA's support for construction, operation,
and science analysis for the {\it GALEX} mission,
developed in cooperation with the Centre National d'Etudes Spatiales
of France and the Korean Ministry of Science and Technology. 

{\it Facilities:} \facility{GALEX}

\clearpage

\begin{deluxetable}{cc}
\tablewidth{0pt}
\tablecaption{Sample selection limits}
\tablehead{ \colhead{Parameter} & \colhead{Limits}}
\startdata
SDSS primary target flag & TARGET\_GALAXY \\
$r$-band magnitude\tablenotemark{a} & $14.5 < r < 17.6$ \\
$r$-band magnitude error & $\sigma_r <  0.2$ \\
redshift & $ 0.01 < z <  0.25$ \\
redshift confidence & $z_{conf} > 0.67$ \\
{\it GALEX} field radius & $fov\_radius < 0\fdg55$ \\
{\it GALEX} exposure time & $t > 1000$ s \\
NUV flag & $nuv\_artifact \leq 1$ \\
FUV and NUV apparent magnitude\tablenotemark{a} & $16.0 < N(F)UV < 23.0$ \\
\enddata
\tablenotetext{a}{The $r$-band magnitude limits were applied after correcting for foreground Galactic extinction while those for the $NUV$ and $FUV$ were applied before the extinction correction.}
\label{selection_limits}
\end{deluxetable}

\clearpage

\begin{deluxetable}{c}
\tablecaption{Galaxy volume densities vs. $(NUV-r)_{0.1}$ and $M_{r,0.1}$
\tablenotemark{a}}
\tablehead{ \colhead{}}
\startdata
\enddata
\tablenotetext{a}{The data for this table are available in the electronic edition of the journal only. The data are in units of Mpc$^{-3}$ mag$^{-2}$.}
\label{cmd_nuv_tab}
\end{deluxetable}

\begin{deluxetable}{c}
\tablecaption{Errors in the galaxy volume densities vs. $(NUV-r)_{0.1}$ and $M_{r,0.1}$
\tablenotemark{a}}
\tablehead{ \colhead{}}
\startdata
\enddata
\tablenotetext{a}{The data for this table are available in the electronic edition of the journal only. The data are in units of Mpc$^{-3}$ mag$^{-2}$ and correspond to the statistical errors only calculated using equation (\ref{phierr_eqn}).}
\label{cmderr_nuv_tab}
\end{deluxetable}

\begin{deluxetable}{c}
\tablecaption{Galaxy volume densities vs. $(FUV-r)_{0.1}$ and $M_{r,0.1}$
\tablenotemark{a}}
\tablehead{ \colhead{}}
\startdata
\enddata
\tablenotetext{a}{The data for this table are available in the electronic edition of the journal only. The data are in units of Mpc$^{-3}$ mag$^{-2}$.}
\label{cmd_fuv_tab}
\end{deluxetable}

\begin{deluxetable}{c}
\tablecaption{Errors in the galaxy volume densities vs. $(FUV-r)_{0.1}$ and $M_{r,0.1}$
\tablenotemark{a}}
\tablehead{ \colhead{}}
\startdata
\enddata
\tablenotetext{a}{The data for this table are available in the electronic edition of the journal only. The data are in units of Mpc$^{-3}$ mag$^{-2}$ and correspond to the statistical errors only calculated using equation (\ref{phierr_eqn}).}
\label{cmderr_fuv_tab}
\end{deluxetable}

\clearpage
\begin{deluxetable}{lllll}
\tablewidth{0pt}
\tablecaption{$r_{0.1}$ Schechter Function Parameters vs. Color}
\tablehead{\colhead{$(NUV-r)_{0.1}$} & \colhead{$\log{\phi^*}$} & \colhead{$M_{r,0.1}$} & \colhead{$\alpha$} & \colhead{$\log{\rho_{r,0.1}}$} \\
 & \colhead{(Mpc$^{-3}$)} & & & \colhead{(ergs s$^{-1}$ Hz$^{-1}$ Mpc$^{-3}$) }
 }
\startdata
0.50 & $-4.603\pm0.960$ & $-20.454\pm0.995$ & $-1.913\pm0.840$ & $24.931\pm0.838$\tablenotemark{a} \\
1.50 & $-2.871\pm0.185$ & $-20.331\pm0.250$ & $-1.465\pm0.175$ & $26.121\pm0.084$ \\
2.50 & $-2.564\pm0.060$ & $-20.786\pm0.105$ & $-0.894\pm0.140$ & $26.368\pm0.032$ \\
3.50 & $-2.775\pm0.045$ & $-20.711\pm0.125$ & $-0.357\pm0.220$ & $26.103\pm0.035$ \\
4.50 & $-2.962\pm0.065$ & $-20.874\pm0.155$ & $-0.579\pm0.235$ & $25.975\pm0.042$ \\
5.50 & $-2.864\pm0.070$ & $-21.158\pm0.155$ & $-0.596\pm0.185$ & $26.187\pm0.039$ \\
6.50 & $-3.474\pm0.125$ & $-21.127\pm0.345$ & $+0.022\pm0.660$ & $25.621\pm0.087$ \\
All colors & \nodata & \nodata & \nodata & $26.898\pm0.022$\tablenotemark{b} \\
\enddata
\tablenotetext{a}{When determining $\rho_{r,0.1}$ for the bluest bin, we have only integrated the luminosity function down to an absolute magnitude $M_{r,0.1}=-12$, rather then to zero luminosity in the other color bins.}
\tablenotetext{b}{The total luminosity density is derived by summing all of the values for $\rho_{r,0.1}$ in each color bin.}
\tablecomments{All the errors quoted in this table are the statistical errors only derived by taking all combinations of parameters within one of the minimum reduced $\chi^2$. All values quoted in this table assume $H_0=70$ km s$^{-1}$ Mpc$^{-3}$, $\Omega_m=0.3$, and $\Omega_{\Lambda}=0.7$.}
\label{schechter_r}
\end{deluxetable}

\begin{deluxetable}{lllll}
\tablewidth{0pt}
\tablecaption{$NUV_{0.1}$ Schechter Function Parameters vs. Color}
\tablehead{\colhead{$(NUV-r)_{0.1}$} & \colhead{$\log{\phi^*}$} & \colhead{$M^*_{NUV,0.1}$} & \colhead{$\alpha$} & \colhead{$\log{\rho_{NUV,0.1}}$} \\
 & \colhead{(Mpc$^{-3}$)} & & & \colhead{(ergs s$^{-1}$ Hz$^{-1}$ Mpc$^{-3}$) }
}
\startdata
0.50 & $-4.604\pm0.900$ & $-19.792\pm0.995$ & $-1.786\pm0.730$ & $24.467\pm0.402$\tablenotemark{a} \\
1.50 & $-2.707\pm0.130$ & $-18.428\pm0.185$ & $-1.284\pm0.130$ & $25.409\pm0.047$ \\
2.50 & $-2.566\pm0.050$ & $-18.341\pm0.090$ & $-0.943\pm0.105$ & $25.397\pm0.024$ \\
3.50 & $-2.862\pm0.045$ & $-17.545\pm0.105$ & $-0.599\pm0.145$ & $24.744\pm0.026$ \\
4.50 & $-3.047\pm0.060$ & $-16.605\pm0.125$ & $-0.746\pm0.150$ & $24.192\pm0.030$ \\
5.50 & $-2.880\pm0.045$ & $-15.689\pm0.095$ & $-0.637\pm0.120$ & $23.985\pm0.026$ \\
6.50 & $-3.469\pm0.080$ & $-15.039\pm0.255$ & $-0.171\pm0.405$ & $23.159\pm0.055$ \\
All colors & \nodata & \nodata & \nodata & $25.791\pm0.029$\tablenotemark{b} \\
\enddata
\tablenotetext{a}{When determining $\rho_{NUV,0.1}$ for the bluest bin, we have only integrated the luminosity function down to an absolute magnitude $M_{NUV,0.1}=-12$, rather then to zero luminosity in the other color bins.}
\tablenotetext{b}{The total luminosity density is derived by summing all of the values for $\rho_{NUV,0.1}$ in each color bin.}
\tablecomments{All the errors quoted in this table are the statistical errors only derived by taking all combinations of parameters within one of the minimum reduced $\chi^2$. All values quoted in this table assume $H_0=70$ km s$^{-1}$ Mpc$^{-3}$, $\Omega_m=0.3$, and $\Omega_{\Lambda}=0.7$.}
\label{schechter_nuv}
\end{deluxetable}

\clearpage

\clearpage 

\clearpage

\clearpage  

\clearpage 

\clearpage 

\clearpage

\clearpage

\clearpage 

\clearpage 

\clearpage

\clearpage

\clearpage  

\clearpage  

\clearpage  

\clearpage 

\clearpage  

\clearpage

\clearpage

\clearpage

\clearpage

\clearpage  

\clearpage

\clearpage 

\clearpage  

\clearpage

\clearpage  

\end{document}